\documentclass[apj]{emulateapj}
\usepackage{apjfonts}
\newcommand{\rmxaa}       {Rev.\ Mexicana Astron.\ Astrofis.}
\usepackage{graphicx}
\newcommand\kms           {km\,s$^{-1}$}

\newcommand{\hii}         {\ion{H}{2}}
\newcommand{\nod}{\nodata}
\journalinfo{The Astrophysical Journal}
\submitted{Received 2007 May 1; accepted 2007 August 8}
\begin{document}
\shorttitle{OH Maser Motions in Star-Forming Regions}
\title{Proper Motions of OH Masers and Magnetic Fields in Massive
Star-Forming Regions}
\author{Vincent L.\ Fish\altaffilmark{1} \& Mark J.\ Reid\altaffilmark{2}}
\shortauthors{Fish \& Reid}
\altaffiltext{1}{Jansky Fellow, National Radio Astronomy Observatory,
  1003 Lopezville Road, Socorro, NM 87801, vfish@nrao.edu}
\altaffiltext{2}{Harvard--Smithsonian Center for Astrophysics, 60
  Garden St., MS 42, Cambridge, MA 02138, reid@cfa.harvard.edu}
\begin{abstract}
We present data of proper motions of OH masers in the massive
star-forming regions ON 1, K3-50, and W51 Main/South.  OH maser
motions in ON 1 are consistent with expansion at approximately
$5$~km\,s$^{-1}$, likely tracing the expanding ultracompact \hii\
region.  Motions in K3-50 are faster and may be indicating the final
stages of OH maser emission in the source, before the OH masers turn
off as the \hii\ region transitions from the ultracompact to the
compact phase.  W51 South shows indications of aspherical expansion,
while motions in W51 Main are more difficult to interpret.
Nevertheless, it appears that the relative projected separation
between W51 Main and W51 South is decreasing, corresponding to an
estimate of enclosed mass of at least $1500~M_\sun$, consistent with
estimates derived from millimeter-wavelength dust emission.  We
confirm the $\sim 20$~mG magnetic fields previously seen in W51 Main,
which may represent the upper end of the density range allowable for
1665~MHz maser emission.  Magnetic field strengths and directions,
obtained from Zeeman splitting, in each source are consistent with
values obtained in the first epoch four to nine years ago.
\end{abstract}

\section{Introduction}
\label{intro}

Maser proper motion observations are a useful tool for understanding
the environment of nascent high-mass stars.  Masers serve as bright,
pointlike tracers of the conditions in the surrounding molecular
material, and detecting proper motions allows the observer to
determine the characteristics of the velocity field on small scales.
Water maser proper motions have been observed for a quarter century
\citep{genzel81a,genzel81b}, and methanol masers are increasingly
being observed for proper motion experiments as well
\citep[e.g.,][]{moscadelli99,minier01}.  Hydroxyl (OH) masers are
common in massive star-forming regions (SFRs), but proper motions of
the ground-state OH transitions at $\lambda = 18$~cm are more
difficult to detect because the angular resolution of a radio array
scales inversely with wavelength.  For this reason, only a few massive
SFRs have observed OH maser proper motions: W3(OH)
\citep{bloemhof92,wright04a} and W75~N \citep{fishw75n} using very
long baseline interferometric (VLBI) techniques, and the nearby
sources Orion \citep{migenes89} and Cepheus~A \citep{migenes92} using
connected-element interferometry.  When available, VLBI observations
are preferred due to the much smaller synthesized beam size, which
allow for more precise positional measurements and reduced blending of
maser spots within the synthesized beam.

Of these sources, the best studied is W3(OH).  \citet{bloemhof92}
clearly established that the OH masers are expanding at
3--5~km\,s$^{-1}$, which is also the expansion rate of the
ultracompact \hii\ region with which they are associated
\citep{kawamura98}.  The maser locations and slow expansion support
the \citet{elitzur78} model in which OH masers form between the
ionization front of an \hii\ region and the shock front ahead of it.
Later observations by \citet{wright04a} confirm the expansion and
detect a rotational component to the motion as well, which is
supported by VLBI observations of the 6.0~GHz excited-state OH masers
in W3(OH) \citep{fishef015}.

This work is motivated by the question of whether expansion, as seen
in W3(OH), is the dominant mode of OH maser motions.  A recent VLBA
survey of the OH masers in massive SFRs provides an adequate starting
point and a first epoch of positional data for a larger sample of
proper motion measurements \citep{fishvlba}.  An analysis of the
proper motions in one source, W75~N, has already been published
\citep{fishw75n}.  In this article we present proper motions
determined from a second epoch of observations of the three sources
with the simplest proper motions: ON~1, K3-50, and W51 Main/South.

It is important to know the distance to a source in order to convert
proper motions, measured in angle units, into velocities.  Distance
measurements toward some sources span a wide range of values.  We have
adopted a distance of 3.0~kpc to ON 1 \citep{araya02}, although values
quoted in the literature range up or down from this value by about a
factor of 2 \citep[e.g.,][]{genzel77,israel83}.  For K3-50 we have
assumed a distance of 8.7~kpc; quoted distances range from less than
8.0 to 9.0~kpc (\citealt{harris75} and references therein).  We assume
a distance to W51 of 7.0~kpc \citep{genzel81b} for consistency with
previous work, although more recent observations of H$_2$O maser
kinematics suggest a distance of $6.1 \pm 1.3$~kpc \citep{imai02},
which is also consistent with this distance.  Since velocities and
lengths (in physical rather than angular units) scale linearly with
distance, these quantities can be easily adjusted to correspond to
better distance determinations in the future.

\section{Observations and Data Analysis}
\label{observations}

\begin{deluxetable*}{llllll}
\tabletypesize{\small}
\tablecaption{Observing Parameters\label{obs-table}}
\tablehead{
  \colhead{} &
  \colhead{} &
  \colhead{} &
  \colhead{} &
  \colhead{Epoch 1} &
  \colhead{Epoch 2} \\
  \colhead{} &
  \colhead{Epoch 1} &
  \colhead{Epoch 2} &
  \colhead{Freq.} &
  \colhead{Beam} &
  \colhead{Beam} \\
  \colhead{Source} &
  \colhead{Date} &
  \colhead{Date} &
  \colhead{(MHz)} &
  \colhead{(mas)} &
  \colhead{(mas)}
}
\startdata
ON~1 & 2000~Nov~22/2001~Jan~06 & 2004~Sep~16/19 & 1665 &   $9.5 \times 6.3$ &  $10.4 \times 7.0$ \\
     &                            &                & 1667 &   $9.5 \times 6.3$ &  $10.6 \times 7.2$ \\
K3-50& 2000~Nov~22/2001~Jan~06 & 2004~Sep~16/19 & 1665 & $39.3 \times 35.8$ & $26.2 \times 22.8$ \\
     &                            &                & 1667 & $39.2 \times 35.7$ & $25.9 \times 22.4$ \\
W51  & 1996~Mar~01/02             & 2005~May~15    & 1665 &  $16.1 \times 9.4$ & $18.7 \times 13.2$ \\
     &                            &                & 1667 &  $16.1 \times 9.4$ & $19.6 \times 13.6$ \\
     &                            &                & 1720 &  \nodata           & $39.7 \times 22.4$
\enddata
\tablecomments{Results from the first epoch are already published in
  \citet{fishvlba}.}
\end{deluxetable*}

Three sources were observed with the VLBA: ON 1 (G69.540$-$0.976),
K3-50 (G70.293$+$1.601), and W51 Main/South (G49.488$-$0.387).  Both
the 1665.40184 and 1667.35903 MHz main-line ($F$-conserving),
ground-state ($^2\Pi_{3/2}, J = 3/2$) transitions of OH were observed.
In the first epoch, ON 1 and K3-50 were observed on two days with
complementary $uv$-coverage (experiment BF064).  Naturally-weighted
images made from the combined data have previously been published by
\citet{fishvlba}.  The observations were taken with 125 kHz bandwidth
divided into 128 spectral channels, for a velocity width of
0.176~\kms\ per channel.  A second epoch of data (BF079) were obtained
with identical observational parameters, with the exception that the
LSR velocity was shifted by 3~\kms\ for all sources due to an error in
observe file preparation.  This was corrected in post-correlation
calibration.  As for the first epoch, the data from the two days of
the second epoch were combined before imaging.  Observing parameters
are summarized in Table~\ref{obs-table}.

In the first epoch of data for W51 (BR039), a 250 kHz bandwidth was
used instead in order to cover the larger velocity range of OH maser
features in this source.  Data for the second epoch (BF085) were taken
using the same bandwidth but with twice the velocity resolution,
providing a velocity width of 0.176~\kms\ per channel.  In the second
epoch, observations of the calibrator J1922+1530 were included for
phase referencing.  A cycle between the calibrator and source
consisted of 80 seconds dwell time on the calibrator and 200 seconds
dwell time on W51.  The 1720.52998~MHz ($F = 2 \rightarrow 1$) line of
OH was also observed in the second epoch.

The data were reduced as described in the appendix of
\citet{fishvlba}.  Images were self-calibrated using the brightest
maser feature in one circular polarization of one transition, and this
calibration was applied to both polarizations.  Additionally,
calibration at 1665~MHz was applied to the 1667~MHz transition.
Therefore, the relative positions of masers in different polarizations
in the 1665 and 1667~MHz transitions within the same epoch are
extremely well known.  The absolute positions of maser features in
each epoch, and therefore the absolute motions of maser features
between epochs, are unknown.  However, we were able to obtain absolute
coordinates for the masers in W51 in the second epoch.

In W51, a bright maser channel was used at 1667~MHz LCP and 1720~MHz
RCP to phase reference the calibrator source J1922+1530.  Periods of
poor phase stability or large fringe rates were flagged, and images of
J1922+1530 were made from the remaining data.  The offset of
J1922+1530 from the position in the VLBA calibrator
list\footnote{\url{http://www.vlba.nrao.edu/astro/calib/vlbaCalib.txt}}
was used to determine the absolute positions of the reference maser
features as well as the relative alignment between the 1665/1667 and
1720~MHz maser data.

Maser spot positions were determined by fitting Gaussians to the
brightest channel of emission for each feature.  An alternate method
of determining maser positions in each epoch is to use a flux-weighted
average of the position across all spectral channels for a maser spot.
However, the small advantage gained from the increased positional
accuracy due to a larger integrated flux is offset by position wander
across the line profile and by inclusion of weaker channels in the
line wings, whose apparent positions are more apt to be affected by
blending with nearby maser spots.  As a test, maser positions in ON 1
and K3-50 were determined in both epochs using both methods.  Within
an epoch, individual maser positions obtained from the two methods
agreed nearly always to better than 1 mas and frequently to better
than 0.2~mas.  Larger deviations were usually due to contamination in
a line wing due to blending with a nearby ($< 2$ beam widths) maser
spot.  Since the brightest-channel method is simpler and less
susceptible to contamination, this method was preferred.  In any case,
proper motion maps produced by the brightest-channel method were
qualitatively similar to those produced by the total-flux method.

Maser motions were identified by aligning the two epochs of data and
finding masers that matched closely in position, velocity, and
dominant circular polarization.  Source-scale motions can introduce
offsets that differ with location in a map but are similar for groups
of maser spots; these motions are of particular interest in this work.
Velocity tolerances were typically about a channel width, with greater
tolerance for particularly broad masers or weak masers whose line
width could not be determined (usually due to lack of detection in
three consecutive channels or spatial blending with another feature).
Maser ``echoes'' (a weak detection at the same position and velocity
but opposite circular polarization as a stronger feature; see
\citet{bf086}) were excluded, as were any spots that could not be
unambiguously identified with a single spot in the other epoch.

\section{Results}
\label{results}

\subsection{ON 1}
\label{results-on1}

\begin{figure}[t]
\resizebox{\hsize}{!}{\includegraphics{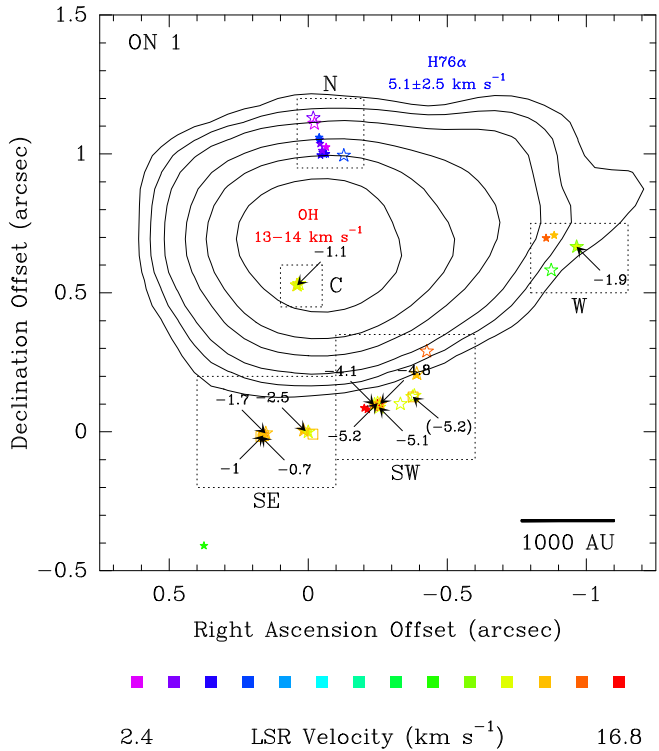}}
\caption{Map of maser emission in ON~1.  Stars denote 1665 MHz masers;
  squares denote 1667 MHz masers.  RCP emission is shown as open
  symbols and LCP as filled symbols.  The color scale indicates LSR
  maser velocities in km\,s$^{-1}$.  Contours indicate 8.4~GHz
  continuum emission from the data in \citet{argon00}.  Coordinates
  are as in \citet{fishvlba}.  Numbers indicate magnetic fields in
  milligauss, with a positive sign indicating that the line-of-sight
  field component is directed away from the Sun.  The value in
  parentheses is derived from assuming that overlapping masers in the
  same polarization but in different transitions are due to incomplete
  Zeeman patters with the same underlying magnetic field strength and
  systemic velocity.  The LSR velocity range of the central cluster of
  OH masers is denoted, as is the velocity of the H76$\alpha$
  recombination line \citep{zheng85}.  The bar indicates 1000~AU at an
  assumed distance of 3.0~kpc.
  \label{on1-map}}
\end{figure}

Detected masers are listed in Table~\ref{on1-table} and shown in
figure \ref{on1-map}.  Our map agrees with that obtained from MERLIN
data by \citet{nammahachak06}.  We find masers near all but one of
their features brighter than 610~mJy, although some of the masers in
the western and southern parts of ON~1 are marginal detections (i.e.,
too weak to be detected above 7\,$\sigma$ in three consecutive
channels) in our data.  We recover 8 of the 11 Zeeman pairs detected
in the first epoch and one component in each of the remaining three
pairs \citep[see Figure 19 of][]{fishvlba}.  In two Zeeman pairs, the
weaker component would be near or below our detection threshold in the
second epoch, assuming the flux density did not change between epochs.
We also find a new Zeeman pair at 1665 MHz ($\Delta$~RA =
$-965.68$~mas RCP and $-964.88$ LCP) and another at 1667 MHz (169.53
RCP and 170.54 LCP) that were not present in our data from the first
epoch.  Magnetic field strengths are consistent to within 0.3 mG
between epochs, with six Zeeman pairs indicating magnetic field
strengths consistent to within 0.1 mG or better.

A reanalysis of data from the first epoch uncovers several spots
missed in \citet{fishvlba}.  The far western region was not imaged;
spots detected in this area in the first epoch reanalysis include an
LCP feature at $(\Delta \alpha [\mathrm{mas}], \Delta \delta
[\mathrm{mas}], v_\mathrm{LSR} [\mathrm{km\,s}^{-1}]) = (-857.56,
697.42, 15.17)$ and two RCP features at $(-966.73, 665.93, 11.82)$ and
$(-874.67, 580.86, 11.65)$, where spatial offsets are in
milliarcseconds from the reference feature in the first epoch and LSR
velocities indicate the velocity of the peak channel of emission in
kilometers per second.  Additionally, a very weak
(76~mJy\,beam$^{-1}$) LCP feature is found in a single channel to the
southeast at $(373.40, -410.27, 11.65)$.  This feature did not meet
our detection criteria since it does not appear in two consecutive
channels, but it does correspond to an identical feature detected in
the second epoch, also very weak and only detected in one channel, and
in the MERLIN observations by \citet{nammahachak06}.

\begin{figure}[t]
\resizebox{\hsize}{!}{\includegraphics{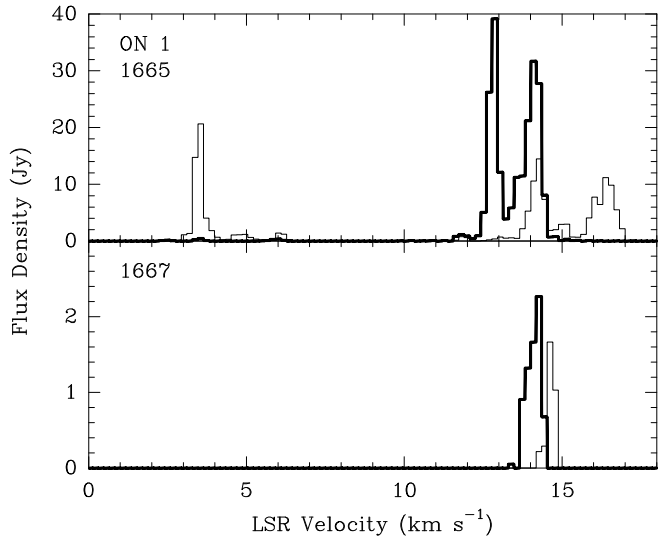}}
\caption{Spectra of recovered maser emission in ON~1.  RCP emission is
shown in bold and LCP in normal weight.  Emission in the 1665 and 1667
MHz transitions is shown in the top and bottom panels,
respectively.\label{on1-spectra}}
\end{figure}

Figure \ref{on1-spectra} shows the recovered maser spectra for ON~1.
In each spectral channel, the fitted maser emission is summed and
displayed.  This is different from total power spectra in that only
emission recovered from fitting is summed; for instance, the recovered
emission in a channel with no detected masers is identically zero.
The spectra are qualitatively similar to those found in
\citet{argon00}, noting their different flux convention and an
erroneous 1~\kms\ shift in their spectra for this source only.
Differences in the spectra are likely due to maser variability,
although the \citet{argon00} spectra may underestimate total flux
density when multiple masing regions are seen at the same LSR velocity
(see their \S 5).

\begin{figure}[t]
\resizebox{\hsize}{!}{\includegraphics{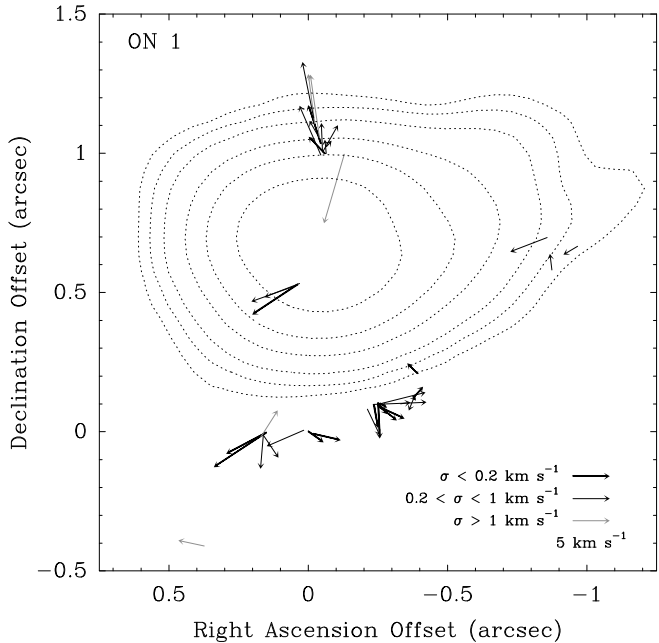}}
\caption{Plot of motions of maser spots in ON 1.  Motions are
  indicated by arrows and correspond to motions in 4 years.  Arrow
  lengths are proportional to velocity as indicated in the lower right
  by arrows representing a motion of 5~km\,s$^{-1}$, assuming a
  distance of 3.0 kpc.  The constant vector has been chosen to
  minimize the total length of all vectors, weighted by the square of
  the effective brightness of each maser spot.
  \label{on1-motions}}
\end{figure}

Figure \ref{on1-motions} shows a map of motions of maser spots
detected in both epochs. Arrow weight and color indicate the random
errors in velocity determination based on the signal-to-noise ratio in
each epoch.  There is also an absolute positional error associated
with the registration of maser positions between epochs.  As described
in \S \ref{observations}, our data are consistent with any map that
differs from that shown by the addition of a single constant vector to
all motions.  Since our data were not phase-referenced, we do not know
the absolute locations of our reference maser feature in each epoch
(which is equivalent to the constant vector of motion of the reference
feature).  In Figure \ref{on1-motions}, the constant vector has been
chosen to minimize the total length of all vectors, weighted by the
square of the effective brightness of each maser spot.  This vector
corresponds to a motion of $-1.7$~\kms\ toward the East and
$-1.5$~\kms\ toward the North, in a frame in which the reference maser
spot at the origin is stationary.  The constant vector is not highly
sensitive to the weighting used; weighting by the first power of the
effective brightness or using constant (equal) weights for all maser
spots results in similar maps to within $\approx 1$~\kms.

In order to minimize confusion, it is useful to clearly define several
terms that will be used henceforth to quantify maser motions.
Consider two maser spots, $a$ and $b$, each detected in epochs 1 and
2.  The separation of the two maser spots is defined to be $Sep_i^{ab}
\equiv \sqrt{(\alpha_i^a-\alpha_i^b)^2 + (\delta_i^a-\delta_i^b)^2}$,
where the subscript $i$ refers to the epoch (1 or 2) and the
superscript identifies the maser spot.  This quantity has units of
angle (milliarcseconds).  The change in separation of the pair of
maser spots is defined to be $Sep_2^{ab} - Sep_1^{ab}$.  The pairwise
expansion velocity (measured in km\,s$^{-1}$) of spots $a$ and $b$ is
$(Sep_2^{ab} - Sep_1^{ab}) D t^{-1}$, where $D$ is the distance to the
source and $t$ is the time between the two epochs of observations.
The pairwise expansion velocity is positive if the change in
separation is positive, i.e., if the two masers have moved apart from
each other.

\begin{figure}[t]
\resizebox{\hsize}{!}{\includegraphics{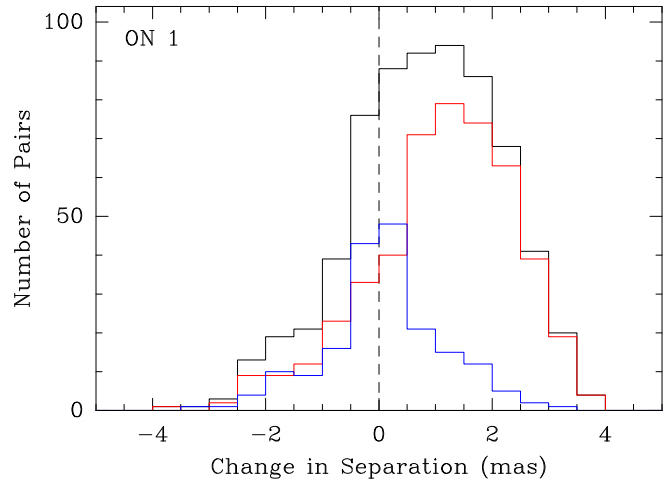}}
\caption{Histogram of the change in separation of pairs of maser spots
  in ON 1.  The solid black histogram shows all possible pairs drawn
  from groups N, C, SW, and SE as defined in Figure~\ref{on1-map}.
  The red histogram shows all pairs such that both maser spots are not
  in the same group.  The blue histogram includes only pairs of maser
  spots within the same group.  The isolated spot spot in the
  southeast and the spots in group W are excluded from this analysis
  because there are no bright masers in either case.  Motions within a
  group appear to show neither expansion nor contraction.  On large
  scales, there is a bias toward positive shifts, indicating expansion
  of about 1.5~mas over 4 years (6 km\,s$^{-1}$ at a distance of 3.0
  kpc).  An error of 1~km\,s$^{-1}$ in determining a maser motion
  would correspond to 0.27~mas.
  \label{on1-seps}}
\end{figure}

\begin{figure}[t]
\resizebox{\hsize}{!}{\includegraphics{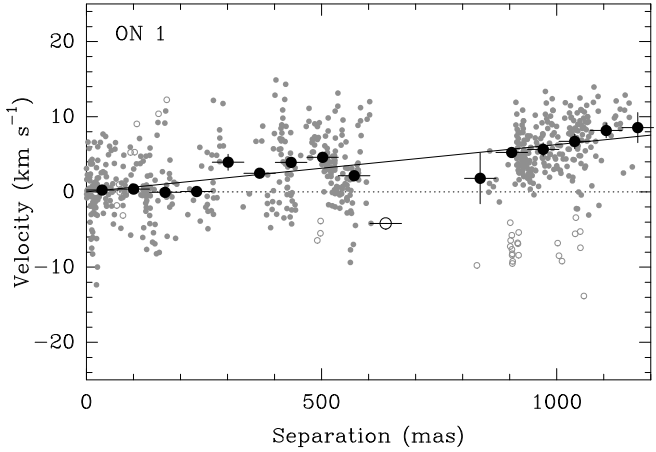}}
\caption{Plot of the relative pairwise expansion velocities of maser
  spots in groups N, C, SW, and SE as defined in Figure~\ref{on1-map}.
  Expansion velocities for pairs of masers are shown as small grey
  circles.  See \S~\ref{results-on1} for definitions of Velocity and
  Separation.  Excluding the maser spot near $(\Delta\alpha,
  \Delta,\delta) \approx (-0\farcs13, 1\farcs0)$ in Figure
  \ref{on1-motions} (small open circles), the line of best fit
  indicates an expansion of $(6.31 \pm 0.23) \times
  10^{-3}$~km\,s$^{-1}$\,mas$^{-1}$ (D = 3.0 kpc).  The vertical
  scatter is much greater than typical random measurement error
  ($\lesssim 1$~km\,s$^{-1}$) and indicates that the random component
  of maser proper motions is large.  Large black circles indicate the
  mean and standard error of the mean for the data in 200~AU bins
  (67~mas at 3.0~kpc), excluding the small open circles; the large
  open circle indicates the value for a bin with exactly one data
  point, for which the standard error of the mean cannot be estimated.
  For definiteness, we have assumed spherical expansion but cannot
  distinguish this from more complex models.  See \S~\ref{discussion}
  for details.
\label{on1-hubble}}
\end{figure}

Figure \ref{on1-seps} shows a histogram of the change in pairwise
separation ($Sep_2^{ab} - Sep_1^{ab}$) of maser spots in ON 1 in
groups N, C, SW, and SE.  As discussed in \citet{bloemhof92}, this
allows a model-independent and non-parametric test for expansion or
contraction.  The distribution is biased toward positive shifts,
indicating that net expansion is occurring.  The expansion is even
more noticeable when pairs containing two maser spots in the same
group are excluded.  The distribution of the change in separation of
pairs containing two maser spots \emph{within} the same group is
centered near zero.  This suggests that motions within a group are
coherent, while expansion dominates on larger scales.  A plot of this
expansion is presented in Figure \ref{on1-hubble}, which shows the
expansion velocity of pairs of maser spots
$(Sep_2^{ab}-Sep_1^{ab})Dt^{-1}$ as a function of their separation
$Sep_1^{ab}$ for pairs of maser spots $a$ and $b$.  The line of best
fit gives an expansion rate of $(6.31 \pm 0.23) \times
10^{-3}$~km\,s$^{-1}$\,mas$^{-1}$ (D = 3.0 kpc), or $(434 \pm
16)$~\kms\,pc$^{-1}$.  Assuming that the motion of the masers near the
southern limb of ON 1 is mostly in the plane of the sky, this suggests
that the expansion velocity of the masers is 4 -- 5 km\,s$^{-1}$,
comparable to a typical maser motion in ON~1 (Fig.~\ref{on1-motions})
as well as the expansion velocity of the OH masers in W3(OH)
\citep{bloemhof92}.

Approximately 83\% of maser motions are consistent with expansion
(i.e., oriented in a semicircle away from the expansion center), with
the remainder consistent with contraction.  However, this is not a
robust statistic, as the percentages depend on the constant vector of
motion, which is not known a priori.

The H76$\alpha$ recombination line velocity of the \hii\ region is
$5.1 \pm 2.5$~km\,s$^{-1}$ \citep{zheng85}.  This seems to suggest
that the southern and central masers in ON 1 are redshifted with
respect to the \hii\ region and the northern masers are slightly
blueshifted (Figure 5).  From the relative LSR velocities of the OH
masers and the \hii\ region, \citet{zheng85} concluded that the OH
masers are still undergoing infall toward the central (proto)star.
Indeed, the LSR velocity of the central group of masers is
13.1--13.9~km\,s$^{-1}$, which is redshifted compared to the central
condensation.  This maser cluster is probably located in front of the
\hii\ region, the core of which may be optically thick in continuum
emission \citep{zheng85}.  (While the northern group and possibly some
masers in the southern group appear to be projected atop the \hii\
region as well, possible registration errors of a few tenths of an
arcsecond between the maser and continuum maps prevent a definitive
determination of their location along the line of sight with respect
to the \hii\ region.)  However, optical depth effects may dominate
determination of the systemic velocity from hydrogen recombination
lines \citep{berulis83,welch87,keto95}.  Indeed, the velocity as
determined from ammonia observations is 11.6~km\,s$^{-1}$
\citep{harju93}.  It is otherwise difficult to reconcile the net
expansion we observe in the maser motions with the contraction that
\citet{zheng85} deduce from their recombination line velocity.  The
difference in LSR velocity between the northern maser group and the
rest of the masers can be explained if, for instance, large-scale
rotation contributes to the maser motions as well.

The scatters in the proper motions in Right Ascension and Declination
($\mu_x, \mu_y$) are $\sigma_{\mu_x} = 0.85$~mas and $\sigma_{\mu_y} =
1.02$~mas, respectively.  At a source distance of 3.0~kpc, these
correspond to a velocity scatter of $\sigma_{v_x} = 3.2$~\kms\ and
$\sigma_{v_y} = 3.8$~\kms.  The scatter in the distribution of the LSR
velocities of the same set of masers is $\sigma_{v_z} = 4.8$~\kms, a
value that does not change when the velocities of the central and
southern clusters of masers are corrected for the effects of Zeeman
splitting (i.e., ``demagnetized'').  This high scatter is almost
entirely due to the fact that ground-state OH maser LSR velocities in
ON~1 appear in disjoint velocity ranges centered near 4~km\,s$^{-1}$
and 14~km\,s$^{-1}$, with no masers at intermediate velocities.

It is not surprising that $\sigma_{v_y} > \sigma_{v_x}$, since the
observed OH maser clumps are preferentially distributed in the
north-south (Declination) direction.  In any case, the uncertainty in
the distance to the source corresponds to a factor of two uncertainty
in $\sigma_{v_x}$ and $\sigma_{v_y}$, precluding any conclusive
determination of whether the expansion velocity of the masers along
the line of sight is significantly different from that in either
transverse direction.

\subsection{K3-50}
\label{results-k350}

\begin{figure}[t]
\resizebox{\hsize}{!}{\includegraphics{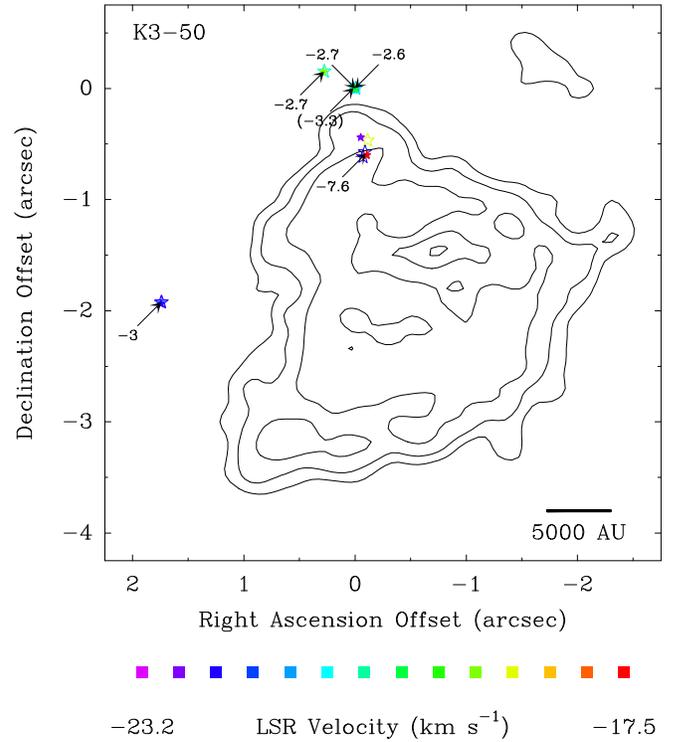}}
\caption{Map of maser emission in K3-50.  The velocity of the
  ultracompact \hii\ region, as derived from H76$\alpha$ observations,
  is nominally $-33 \pm 1$~km\,s$^{-1}$ but spans a range of
  $-54$~km\,s$^{-1}$ in the southeast to $+2$~km\,s$^{-1}$ in the
  northwest \citep{depree94}, complicating comparison with OH maser
  velocities.  The NH$_3$ velocity is $-26$~km\,s$^{-1}$
  \citep{churchwell90}.  Symbols are as in Figure \ref{on1-map}.
  \label{k350-map}}
\end{figure}

\begin{figure}[t]
\resizebox{\hsize}{!}{\includegraphics{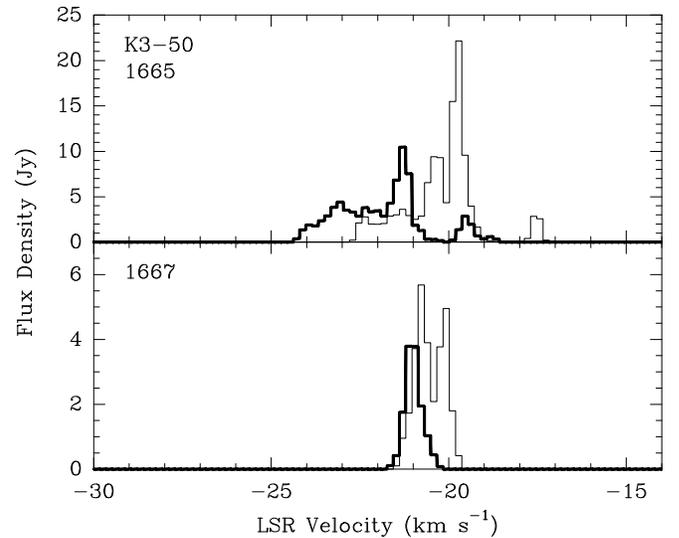}}
\caption{Spectra of recovered maser emission in K3-50.  See Figure
  \ref{on1-spectra} for details.
  \label{k350-spectra}}
\end{figure}

Detected masers in K3-50 are listed in Table~\ref{k350-table} and
shown in Figure~\ref{k350-map}.  The magnetic fields derived from
Zeeman splitting have not changed since the first epoch of
observations.  We recover five of the six Zeeman pairs detected in the
first epoch \citep[see Figure 21 of][]{fishvlba}.  The magnetic field
strengths obtained in the second epoch agree with those obtained in
the first epoch to 0.1 mG or better.  This is within the errors in
determining the central velocity of the $\sigma$-components of a
Zeeman pair from our data.  Magnetic field values are in excellent
agreement with Zeeman pairs obtained from VLA observations in 1993
\citep{argon00,fish03}.  The recovered maser spectrum
(Fig.~\ref{k350-spectra}) also closely matches the VLA spectrum of
\citet{argon00}.

\begin{figure}[t]
\resizebox{\hsize}{!}{\includegraphics{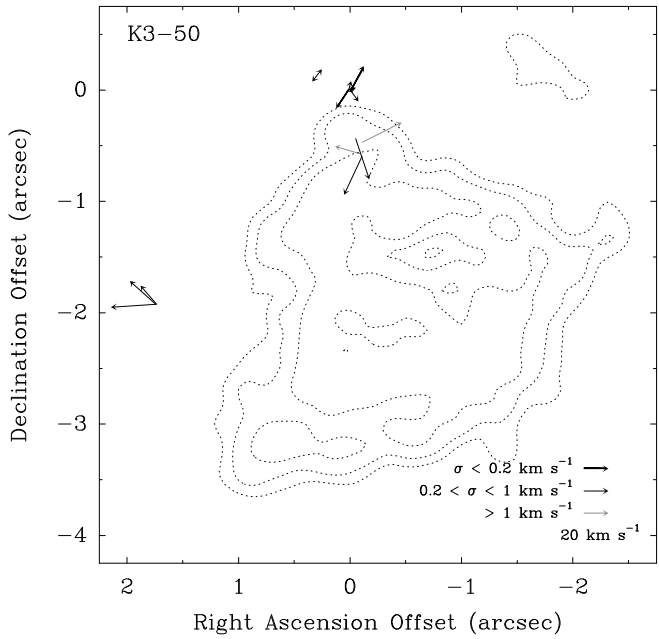}}
\caption{Plot of motions of maser spots in K3-50.  See Figure
  \ref{on1-motions} for details.  The continuum source is K3-50~A.
  Weaker continuum emission extends to the northwest and southeast, as
  seen in the 14.7~GHz maps of \citet{depree94}.\label{k350-motions}}
\end{figure}

\begin{figure}[t]
\resizebox{\hsize}{!}{\includegraphics{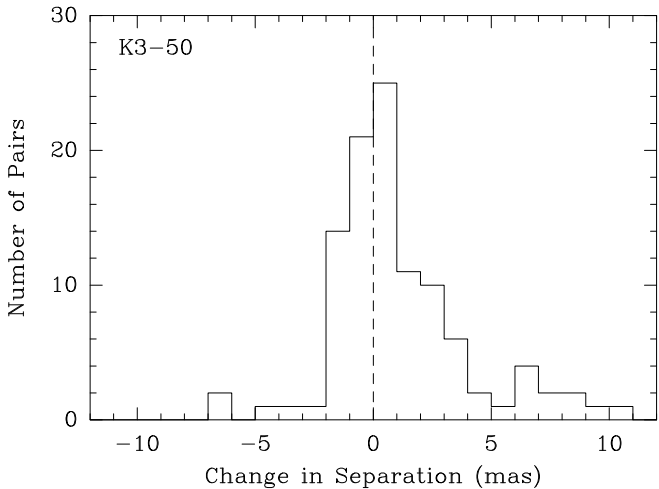}}
\caption{Histogram of the change in separation of pairs of maser spots
  in K3-50.  As with ON 1, the histogram is biased toward positive
  shifts, consistent with expansion.  An error of 1~km\,s$^{-1}$ in
  determining a maser motion would correspond to
  0.10~mas.\label{k350-seps}}
\end{figure}

Proper motions of the OH masers in K3-50 are shown in Figure
\ref{k350-motions}.  As for ON~1, we have chosen the constant vector
to minimize the total of proper motion vector lengths, weighted by the
square of the effective brightness.  Figure \ref{k350-seps} shows a
histogram of the change in pairwise separation of maser spots in
K3-50.  As with ON 1, there is a general trend for maser separations
to increase with time, indicative of expansion, although the magnitude
of the motions is larger.  About 60\% of the motions are consistent
with expansion (with the large caveat noted in \S~\ref{results-on1}).

A major limitation of the analysis is that masers are found only at or
near the northern limb of the \hii\ region and offset to the east of
the \hii\ region.  With such small areal coverage, it is difficult to
draw conclusions as to the large-scale motions around the \hii\
region.  Nevertheless, it does appear that the motion of the eastern
clump of masers is directed away from the \hii\ region.

The environment of K3-50~A is complicated and energetic.  From
continuum and radio recombination line data, \citet{depree94} report a
high-velocity bipolar outflow at position angle $160\degr$ with a
velocity gradient of 6~\kms~arcsec$^{-1}$ across K3-50~A increasing
from the southeast to the northwest.  \citet{hofmann04} find evidence
of at least ten stars in K3-50~A, including at least seven in the
central square arcsecond.  Mid-infrared imaging indicates that there
may be three ionizing sources in the \hii\ region, although none are
coincident with the OH maser locations \citep{okamoto03}.

\subsection{W51}
\label{results-w51}

\begin{figure}[t]
\resizebox{\hsize}{!}{\includegraphics{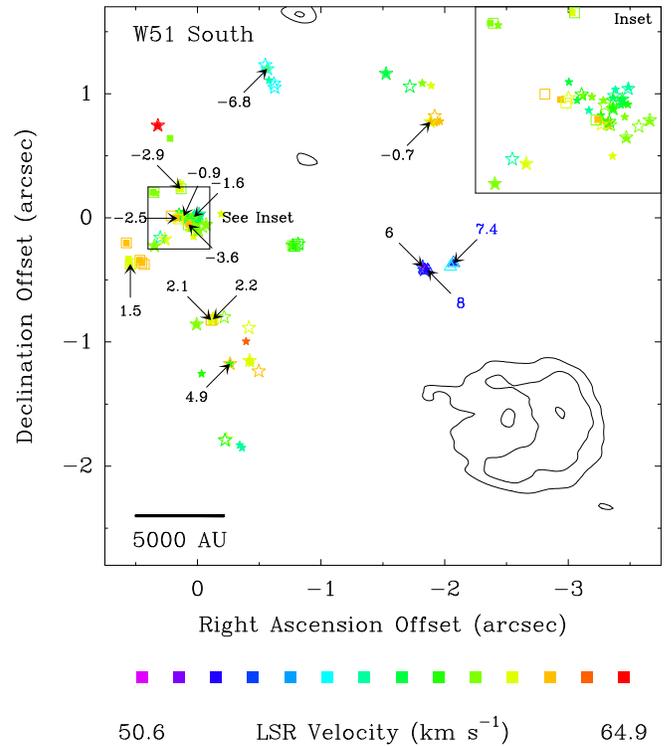}}
\caption{Map of maser emission in W51 South.  Triangles denote
  1720~MHz masers.  The position of the reference feature at the
  origin in the second epoch is $19^\mathrm{h}23^\mathrm{m}43\fs9715,
  14\degr30\arcmin28\farcs395 \pm 0\farcs01$ (J2000).  The H92$\alpha$
  recombination line velocity obtained for source e1 (continuum in the
  southwest) is $+55$~km\,s$^{-1}$ \citep{mehringer94}.  Symbols are
  as in Figure \ref{on1-map}.  Magnetic fields derived from 1720~MHz
  Zeeman pairs are indicated in blue.  An inset is included to show
  the cluster near the origin more clearly.
  \label{w51s-map}}
\end{figure}

\begin{figure}[t]
\resizebox{\hsize}{!}{\includegraphics{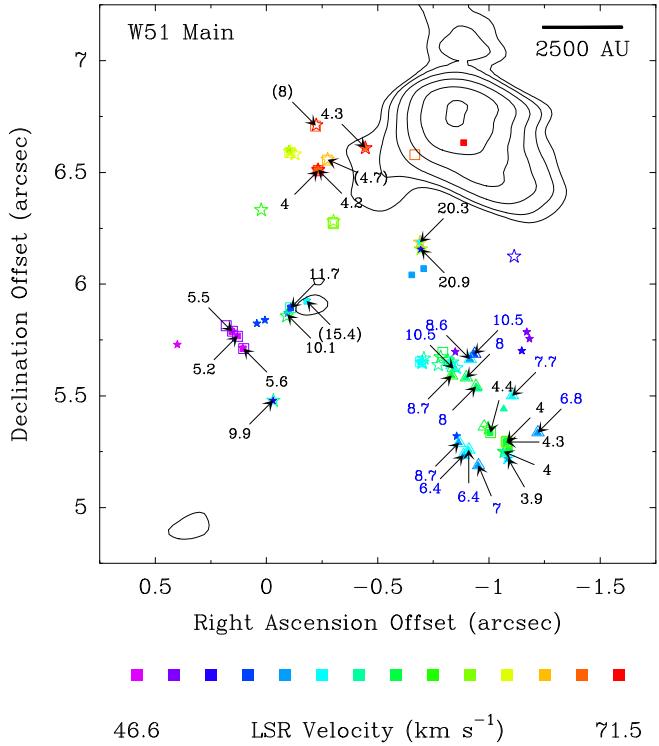}}
\caption{Map of maser emission in W51 Main.  Coordinates are offsets
  from the reference feature at the origin in W51 South.  No
  H92$\alpha$ recombination line emission is detected toward the
  continuum source e2 \citep{mehringer94}. See Figure \ref{on1-map}
  for details.
  \label{w51m-map}}
\end{figure}

Detected maser spots in W51 South and Main are listed in
Tables~\ref{w51s-table} and \ref{w51m-table} and shown in Figures
\ref{w51s-map} and \ref{w51m-map}, respectively.  The relative
positions of the 1665/1667~MHz and 1720~MHz frames were determined by
analysis of the offsets from the calibrator source J1922+1530.  The
brightest maser in 1667~MHz LCP was used to phase reference the
1665/1667~MHz data because its spectral channel is free from other
significant emission.  The brightest maser at 1720~MHz RCP was used to
determine the absolute positions of the 1720~MHz data.  Zero right
ascension and declination offsets in Tables~\ref{w51s-table} and
\ref{w51m-table} and Figures~\ref{w51s-map} and \ref{w51m-map}
corresponds to $19^\mathrm{h}23^\mathrm{m}43\fs9715,
14\degr30\arcmin28\farcs395$ (J2000).  Formal errors in determining
the positions of the calibrator and maser spot are on the order of
1~mas, although ionospheric effects and possible source structure in
the calibrator likely limit accuracy to approximately 10~mas.

\begin{figure}[t]
\resizebox{\hsize}{!}{\includegraphics{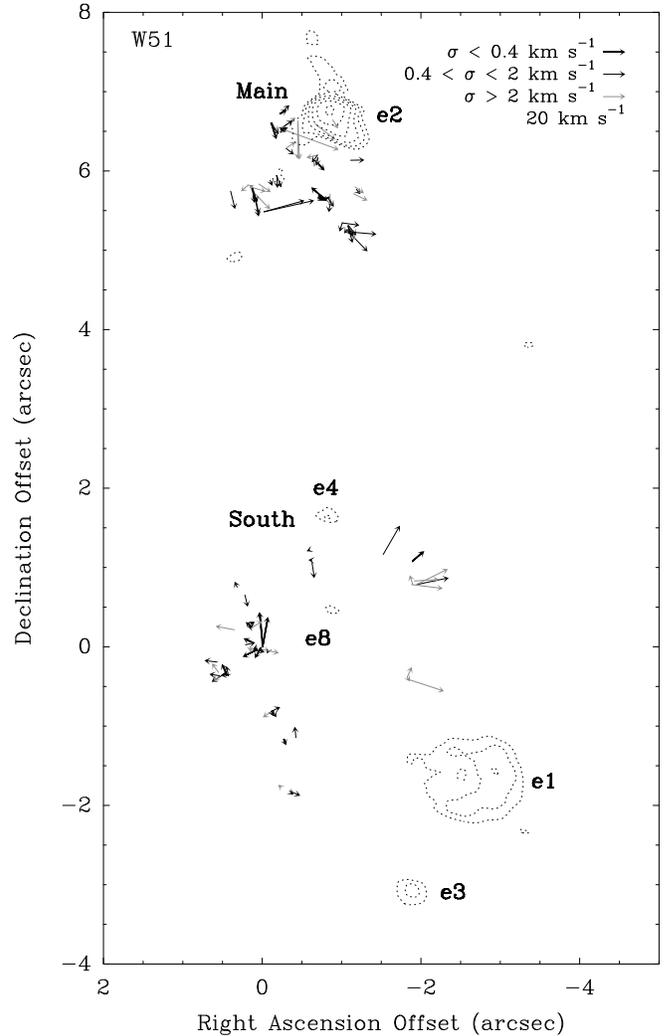}}
\caption{Plot of motions of maser spots in W51 Main/South.  The
  contours indicate 22.4 GHz continuum emission, taken from the VLA
  archive (project code AK560).  We estimate that the possible
  alignment error between the maser data and the continuum emission is
  several tenths of an arcsecond.  The reference velocity frame has
  been chosen to minimize the total length of vectors in W51 South
  only.
  \label{fig-w51}}
\end{figure}

The relative alignment of the 1720~MHz data with respect to the
main-line masers closely matches that of \citet{gaume87} and is
broadly consistent with the alignment determined by \citet{benson84}.
Our data indicate that the 1720~MHz masers are likely located several
hundred milliarcseconds south of their locations in \citet{argon00}
relative to the main-line masers.  Our positions are also slightly
south of the coordinates obtained by \citet{gaume87} after epoch
precession.

The distribution of masers at 1720~MHz is unlike the distribution of
main-line masers.  Only six relatively weak (brightest
2.19~Jy\,beam$^{-1}$) 1720~MHz masers are seen in W51 South, all in
the west of the source.  Magnetic field values are similar to, though
slightly higher than, the magnetic field value obtained from the
1665~MHz Zeeman pair in this cluster.  In contrast, there are numerous
strong 1720~MHz masers in W51 Main, all located in the southwestern
group of maser spots, with six masers brighter than
10~Jy\,beam$^{-1}$.  All 12 1720~MHz Zeeman pairs indicate a stronger
magnetic field than any of the 5 main-line Zeeman pairs in the region.
We have used a Zeeman splitting coefficient of 0.654~kHz\,mG$^{-1}$ to
obtain magnetic field strengths at 1720~MHz since only $\sigma^{\pm
1}$ components produce detectable masers \citep{bf086}.

The large ($\sim 20$~mG) magnetic fields seen in the 1665~MHz masers
in W51 Main \citep[see Figure 18 of][]{fishvlba} persist in the second
epoch.  Small changes in the magnetic field strength estimates of
these two Zeeman pairs ($|\Delta B| = 0.5$ and $0.2$~mG) may not be
significant due to the coarse velocity resolution used in the first
epoch.  We also identify a second cluster of maser spots in which the
magnetic field strength exceeds 10~mG.  The magnetic fields we derive
in W51 Main and W51 South from the second epoch of data are
qualitatively similar to those obtained from the first epoch.  Zeeman
pairings are occasionally ambiguous in the crowded cluster near the
origin, but the pairs we identify agree in sign and approximate
magnitude with features in the first epoch.

The W51 South region contains several continuum sources, including e1,
e4, and e8 \citep[discovered respectively
by][]{scott78,gaume93,zhang98}.  The strongest continuum source, e1,
is offset from the maser emission to the southwest.  Source e4 is
located north of all masers in W51 South.  The only 6035 MHz OH masers
detected with VLBI techniques in W51 are a $+3.6$~mG Zeeman pair
located near source e4 \citep{desmurs98}.  Their total-power spectrum
is suggestive of the existence of several other 6035 MHz masers,
although it is unclear whether these would be associated with W51 d,
W51 Main, or W51 South \citep[see also][]{baudry97}.  Source e8 is
located approximately 2\arcsec\ northeast of source e1, near the
center of the distribution of OH maser spots in W51 South.  Based on
their detection of source e8 at $\lambda = 1.3$~cm but not at $\lambda
= 3.6$~cm, \citet{zhang97} speculate that the continuum flux is
dominated by dust emission heated from a newly-formed massive star, a
conclusion supported by its high flux density at $\lambda = 1.3$~mm
\citep{lai01} compared to centimeter-wave values.  A very rare ammonia
(NH$_3$) $(J,K) = (9,6)$ maser is also seen near source e8
\citep{pratap91}.  This indicates that the e8 region is very
energetic, since it is believed that ortho-NH$_3$ masers are pumped
via $v = 1$ vibrationally-excited states
\citep{mauersberger88,brown91}, located about 1000~K above the ground
$v = 0$ states.

\begin{figure}[t]
\resizebox{\hsize}{!}{\includegraphics{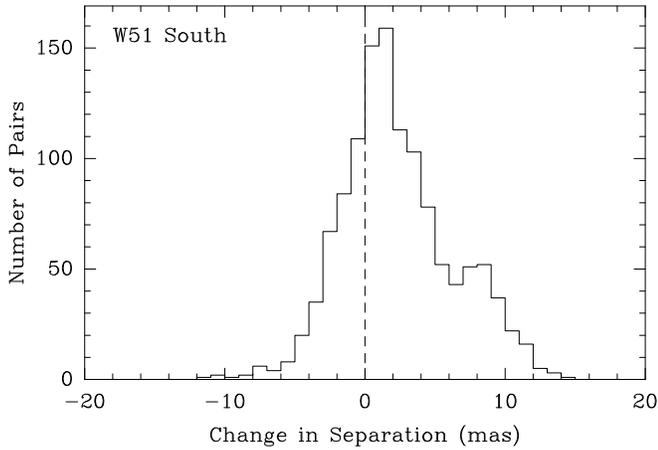}}
\caption{Histogram of the change in separation of pairs of maser spots
  in W51 South.  The bias toward positive shifts is consistent with
  expansion.  An error of 2~km\,s$^{-1}$ in determining a maser motion
  would correspond to 0.28~mas.
  \label{e1-seps}}
\end{figure}

\begin{figure}[t]
\resizebox{\hsize}{!}{\includegraphics{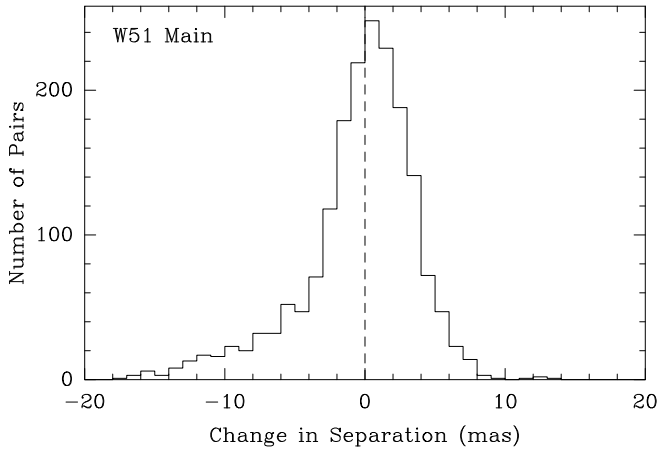}}
\caption{Histogram of the change in separation of pairs of maser spots
  in W51 Main.  The distribution is centered near zero with a larger
  tail toward negative changes in separation.  An error of
  2~km\,s$^{-1}$ in determining a maser motion would correspond to
  0.55~mas.
  \label{e2-seps}}
\end{figure}

Main-line OH maser motions are shown in Figure \ref{fig-w51}.  The OH
maser motions in W51 South are strongly suggestive of expansion
(Figure \ref{e1-seps}), possibly of a shell centered near source e8.
This contrasts with NH$_3$ observations of source e1, which indicate
radial contraction \citep{ho96}.  \citet{zhang98} claim that the
material around the e8 core is contracting at a speed of 3.5~\kms\
based on line asymmetries in CS and CH$_3$CN, although sources e1 and
e8 are barely resolved by their beam.  They also find evidence of
rotation in the e8 core.  To complicate matters further,
\citet{imai02} find evidence of a bipolar outflow traced by H$_2$O
masers in W51 South but find that the source of the outflow is offset
from sources e1, e4, and e8.  \citet{phillips05} also find a 6668 MHz
methanol maser located approximately halfway between sources e1 and
e8.

The W51 Main region is associated with source e2, although most OH
masers are detected to the east and south of this region.  Source e2
appears to be contracting and rotating
\citep{ho96,zhang97,zhang98,sollins04}.  While source e2 is the only
identified continuum source in W51 Main, it is offset from the NH$_3$
emission and OH masers located 1\arcsec\ to the northeast and the
H$_2$O masers located 2\arcsec\ to the north \citep{ho83,gaume93}.
W51 Main also hosts an NH$_3$ (9,6) maser
\citep{madden86,wilson88,pratap91}.  As to the OH masers, the
distribution of the change of separations peaks near zero but has a
larger tail toward negative changes, nominally suggestive of
contraction (Figure \ref{e2-seps}).  Nevertheless, it is difficult to
reconcile these two features of the distribution in terms of a single
kinematic mode.  It is probable that there exist several massive stars
or protostellar condensations driving the masers
\citep[see][]{imai02}, and the disparity between the distributions of
the 1720~MHz and main-line masers suggests that physical conditions
vary significantly in W51 Main.

\begin{figure}[t]
\resizebox{\hsize}{!}{\includegraphics{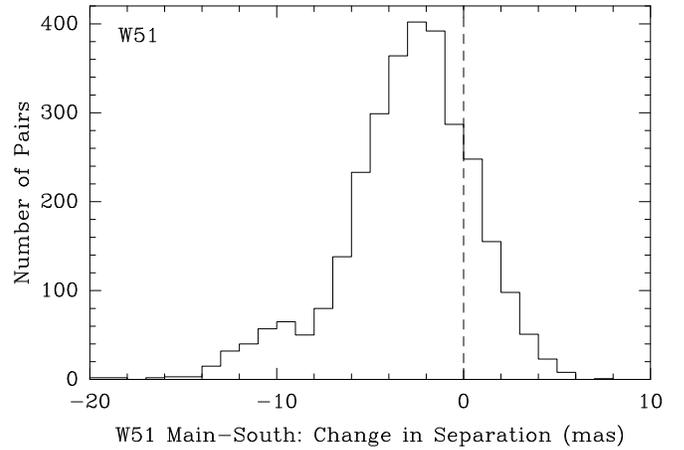}}
\caption{Histogram of the change in separation of pairs of masers with
  one spot in W51 Main and one in W51 South.  A change in separation
  of 1~mas corresponds to a transverse velocity of 3.6~\kms.
  \label{e1e2-seps}}
\end{figure}

It appears that the relative separation between W51 Main and W51 South
is decreasing.  Figure \ref{e1e2-seps} shows the change in separations
between pairs of maser spots such that one spot is in W51 Main and the
other is in W51 South.  The distribution is peaked at less than zero.
This can also be seen in Figure \ref{fig-w51}, in which the reference
frame has been chosen to minimize the total motion in W51 South.  In
this frame, the motions of the masers in W51 Main are clearly biased
toward W51 South.

\begin{figure}[t]
\resizebox{\hsize}{!}{\includegraphics{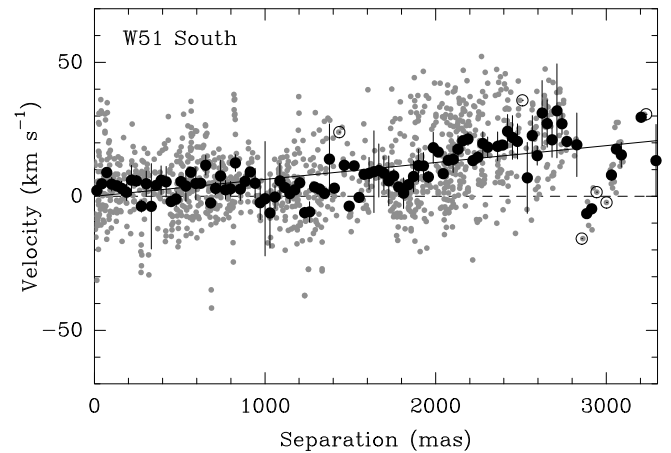}}
\caption{Plot of the relative pairwise expansion velocities of maser
  spots in W51 South.  Symbols are defined as in
  Figure~\ref{on1-hubble}.  See \S~\ref{results-on1} for definitions
  of Velocity and Separation.  The line of best fit indicates an
  expansion of $(6.30 \pm 0.24) \times
  10^{-3}$~km\,s$^{-1}$\,mas$^{-1}$ (D = 7.0 kpc).
\label{w51s-hubble}}
\end{figure}

\begin{figure*}[t]
\resizebox{3.5truein}{!}{\includegraphics{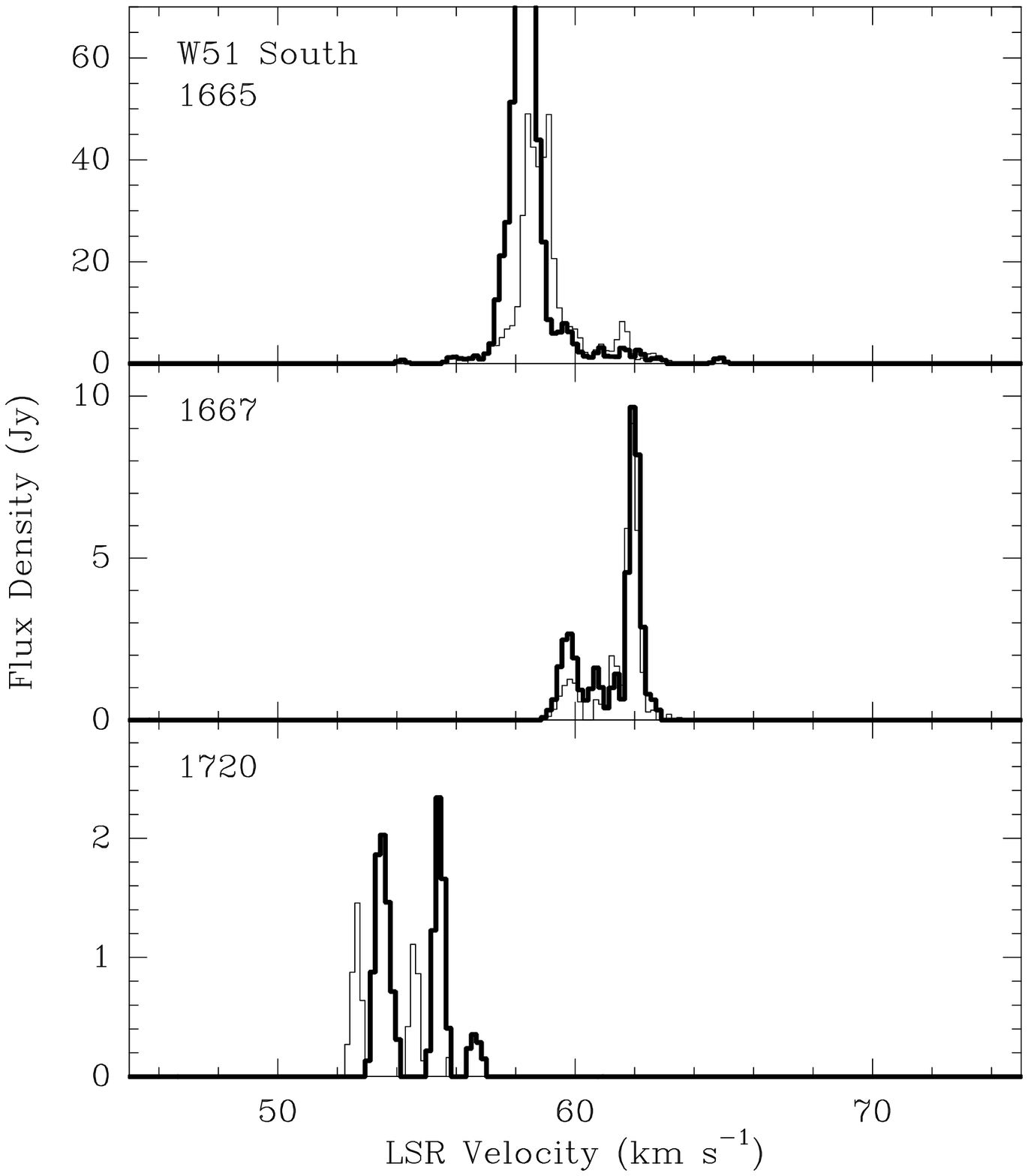}}
\resizebox{3.5truein}{!}{\includegraphics{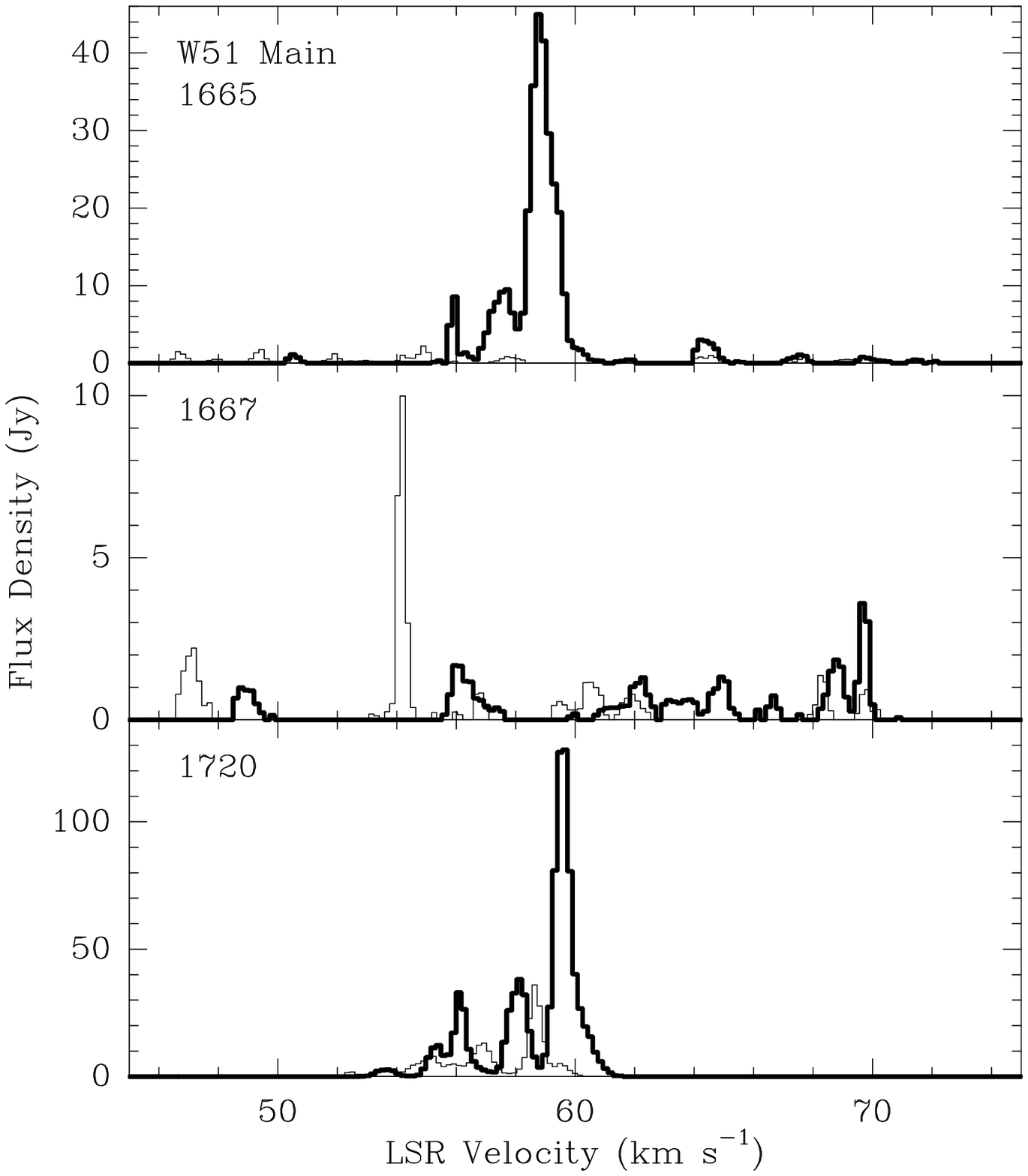}}
\caption{Spectra of recovered maser emission in W51.  Emission is
  shown for W51 South (left) and W51 Main (right).  See Figure
  \ref{on1-spectra} for details.  RCP 1665~MHz emission, which peaks
  near 200~Jy in W51 South, has been clipped for clarity.
  \label{w51-spectra}}
\end{figure*}

The plot of the relative pairwise expansion velocities of the masers
in W51 South is shown in Figure \ref{w51s-hubble}.  Interestingly, the
proportionality constant between radius and expansion velocity is
almost exactly the same as for ON 1.  But the shell of masers in W51
South is larger than in ON 1, and the expansion velocities are
correspondingly higher.  The line of best fit corresponds to a
velocity gradient of $(186~\pm~7)$~\kms\,pc$^{-1}$.  About two-thirds
of maser motions are consistent with expansion (with the caveat noted
in \S~\ref{results-on1}).

The recovered maser emission in W51 is shown in Figure
\ref{w51-spectra}.  As with ON~1 and K3-50, the recovered maser
spectra are qualitatively similar to those of \citet{argon00}.  The
range of maser LSR velocities is much greater in W51 Main than in W51
South, especially at 1667 MHz.  The scatter in the distribution of the
LSR velocities of the masers in W51 South is only $\sigma_{v_z} =
2.7$~\kms\ in W51 South, as compared to $7.5$~\kms\ in W51 Main.
Although the magnetic fields in W51 Main are larger than in the
typical interstellar OH maser source, $\sigma_{v_z}$ drops only to
$7.0$~\kms\ when maser velocities are demagnetized, indicating that
the LSR velocity scatter is due almost exclusively to dynamical
(rather than magnetic) effects.

In contrast, the $1\,\sigma$ scatter in changes in position of the
masers in W51 South and W51 Main are almost identical.  In W51 South
the scatter in the change in Right Ascension is $\sigma_{\mu_x} =
3.0$~mas and the scatter in the change in Declination is
$\sigma_{\mu_y} = 2.5$~mas; in W51 Main, $\sigma_{\mu_x} = 3.1$~mas and
$\sigma_{\mu_y} = 2.5$~mas.  At a distance of 7.0~kpc, this
corresponds to $\sigma_{v_x} = 10.9$~\kms\ and $\sigma_{v_y} =
9.0$~\kms\ for both W51 South and W51 Main.  In W51 Main,
$\sigma_{v_z} \leq \sigma_{v_x}, \sigma_{v_y}$, an effect noted in
W3(OH) as well \citep{bloemhof92}.  But in W51 South, $\sigma_{v_z}
\ll \sigma_{v_x}, \sigma_{v_y}$.  It is probable that the expansion of
the masers in W51 South is not spherical.  For example, \cite{imai02}
note from water maser proper motions the possibility of a bipolar
outflow in W51 South.  An expanding shell model with maser emission
only from the limbs could also produce $\sigma_{v_z} \ll \sigma_{v_x},
\sigma_{v_y}$, although the LSR velocities of the masers in W51~South,
especially the blueshifted 1720~MHz masers, argue against this
interpretation.

\section{Discussion}
\label{discussion}

\subsection{Proper Motions}

A matter of great importance is whether observed maser proper motions
are due to motions of the masing material or due to nonkinematic
effects, such as travelling excitation waves or random coherence in a
turbulent medium.  In the former case, masers can be used as point
tracers of the motions of the material surrounding the central
condensation of a massive SFR.  But if masers arise from travelling
excitation phenomena, their apparent motions do not necessarily
correlate with material motions.

Several lines of evidence support the kinematic interpretation of
maser proper motions.  First, proper motions and radial velocities
have been used to estimate source distances ($D$) from internal
dynamics with great success \citep[e.g.,][]{genzel81b}.  This requires
that, on average, $\mu_x \times D = \mu_y \times D = v_z$, indicating
that motions of brightness peaks are intrinsically related to Doppler
shifts.  Second, proper motions of OH masers in W3(OH) show an
organized, large-scale expansion and rotation
\citep{bloemhof92,wright04a}.  If masers are merely chance lines of
velocity coherence in a turbulent medium, large-scale organization
would not be expected.  Third, maser spots in W3(OH) are observed to
preserve their shape over time, suggesting that the masers are
physically distinct regions of material
\citep{bloemhof96,moscadelli02}.  Fourth, measurements of
excited-state OH absorption in several massive star-forming regions
suggest that the magnetic field of gas traced by OH masers is stronger
than that traced by absorption \citep{fish05,fish07}.  Since the
density $n$ scales approximately as $B^2$
\citep[e.g.,][]{fiebig89,crutcher99}, this implies that masers are
denser than the surrounding, non-masing gas.  This in turn suggests
that masers are physically distinct entities whose observed motions
are real material motions.

In both ON~1 and K3-50, the pairwise separation between maser spots
is increasing.  The same phenomenon is seen in W3(OH)
\citep{bloemhof92,wright04a}.  All of these sources have mature
ultracompact \hii\ regions, which are probably in expansion.  Indeed,
the expansion of the UC\hii\ region in W3(OH) has been detected
directly from multiepoch observations of the continuum emission
\citep{kawamura98}, and the expansion speed is comparable to that
determined from OH maser proper motions.  This is consistent with
models in which OH masers are located in the heated neutral gas
between the ionization and shock fronts of an \hii\ region
\citep[][]{baldwin73,elitzur78}, a picture supported by multiple lines
of observational evidence \citep[e.g.,][]{fish06}.  As the \hii\
region expands, the masers expand with it.

Maser proper motions in K3-50 appear to be faster than in ON 1, even
accounting for the large uncertainty in the distance to the latter.
The continuum source K3-50A is much larger than the \hii\ region in ON
1, indicating that K3-50 is more evolved.  In fact, the diameter of
the continuum region is approximately 0.15~pc, which is comparable to
the size of a hot ammonia core
\citep{vogel87,huttemeister93,depree95}.  This is approximately the
size limit (0.1--0.15~pc) beyond which \emph{ultracompact} \hii\
regions, which often have OH masers, become \emph{compact} \hii\
regions, which do not \citep{habing79}.  The proper motions of the
masers in K3-50 are approximately twice the speed of sound in a
typical \hii\ region ($\sim 13$~km\,s$^{-1}$; \citealt{shu92}),
indicating that the ionization front may be nearly R-critical.  As
this front continues expanding outward into a less dense medium, it
will accelerate, transitioning from a D-type shock (with shocked
neutral gas between the ionization and shock fronts) to an R-type
shock (in which the ionization and shock fronts are coincident),
thereby destroying the OH masers.  It is possible that this has
already happened on the south and west sides of the \hii\ region
K3-50A, where no OH masers are found.  Of course, we cannot rule out
that the masers in K3-50 may be powered by younger, undetected
hypercompact \hii\ regions instead.

In the expansion velocity plots of ON~1 and W51 South (Figures
\ref{on1-hubble} and \ref{w51s-hubble}), OH maser proper motions imply
a velocity gradient on the order of several hundred \kms\,pc$^{-1}$.
Relative motions of the northern and eastern clusters of masers in
K3-50 imply a velocity gradient of similar magnitude.  Gradients in
LSR velocity in ON~1 and W51~Main are also several hundred
\kms\,pc$^{-1}$.  For comparison, the escape velocity is $v_e =
0.094$~\kms~$\sqrt{M/R}$, where $M$ is the enclosed mass in solar
masses and $R$ is the radius of the region in parsecs.  For reasonable
values in a massive star-forming region \citep[$\sim 150~M_\sun$ in
0.06~pc for W51 e2][]{zhang97} the expected velocity gradient is $\sim
80$~\kms\,pc$^{-1}$ for $v = v_e$.  On larger scales, the velocity
gradient $dv/dr$ decreases provided that the enclosed mass grows more
slowly than $r^3$, while the velocity gradient increases on smaller
scales.

The relative proper motion of W51 Main and South, $\sim 8$~\kms\ at a
projected separation of $0.2$~pc, yields a minimum enclosed mass
estimate of $1500~M_\sun$, assuming that the two sources are
gravitationally bound.  This is in line with estimates from
millimeter-wavelength dust emission.  \citet{rudolph90} obtain mass
estimates of $1200$ and $4500~M_\sun$ for W51 e1 and e2, respectively,
from their dust emission at $\lambda = 3.4$~mm, assuming clump
temperatures of 100~K and a dust emissivity $\beta$ of 1.5 (emission
$\propto \lambda^{-\beta}$) based on millimeter and submillimeter
observations of W51 e2.  Interestingly, the 1.3~mm flux density of the
e1/e2 region \citep[20.8~Jy from dust,][]{lai01} corresponds to a
minimum ($\beta = 1$) mass estimate of $1500~M_\sun$, using the
formulation of \citet{hildebrand83} and \citet{ward-thompson95} and
the dust absorption cross section of \citet{beckwith90}.  These
results suggest that maser proper motion measurements may in some
cases provide confirming mass estimates independent of those obtained
from (sub)millimeter dust emission.

It is unclear what the expected a priori functional form of the
expansion velocity plots should be.  For organized expansion, the
relative pairwise expansion velocity should increase as the separation
between the two masers increases, as is observed.  The functional form
depends on such factors as the geometry and thickness of the maser
region as well as the acceleration (or deceleration) of the expansion.
The large scatter of points on these plots prevents a detailed
analysis of these expansion parameters.  Even on the largest scales,
the scatter in pairwise expansion velocities is comparable to the
regular component, indicating that the motions of masers in SFRs are
affected by a large random component in addition to an organized
kinematic mode such as expansion.  Random motions are seen to be quite
large in comparison with maser proper motions, as is noted in water
maser motions as well in some sources, including W51~Main
\citep[e.g.,][]{genzel81b,gwinn94,imai02}.  Despite these limitations,
it is clear that the data indicate a component of expansion, rather
than contraction, and that this expansion can only be seen clearly on
large scales.  Notably, the data do not show expansion for pairwise
separations on the order of $10^{15}$~cm, the clustering scale of OH
masers in star-forming regions \citep{reid80,fish06}.  If expanding at
a typical neutral sound speed of $\sim 1$~km\,s$^{-1}$, a cluster
would double in size in less than 1000 years.  The data suggest that
the size of an OH maser cluster remains fairly stable over time.

\subsection{Magnetic Fields}

Magnetic field strengths imply reasonable densities for OH maser
production.  In molecular clouds, the magnetic field strength $|B|$ is
seen to vary with H$_2$ number density $n$ as $|B| \propto n^\kappa$,
where $\kappa = 0.47 \pm 0.08$ \citep{crutcher99}.  Taking a
line-of-sight magnetic field strength $B_{los} \approx 1$~mG at $n =
10^6$~cm$^{-3}$, typical full magnetic field strengths of less than
about 5~mG correspond to densities less than $10^7$~cm$^{-3}$ assuming
random magnetic field orientations (in which case $B_{los} =
|B|/\sqrt{3}$ statistically).  This density is comfortably within the
range in which ground-state OH masers are predicted to appear
\citep[e.g.,][]{pavlakis96,cragg02}.  Our largest magnetic field
strength of 20.9~mG would thus be expected to correspond to a density
of several times $10^8$~cm$^{-3}$, with an uncertainty of a factor of
several due to the uncertainty in $\kappa$ as well as the natural
scatter of data points as seen in Figure 1 of \citet{crutcher99}.  The
highest density that can produce a detectable 1665~MHz maser depends
on other parameters (such as gas and dust temperatures) but is
approximately $2 \times 10^8$~cm$^{-3}$ in the \citet{cragg02} models.
The scarcity of OH masers indicating magnetic fields of 20~mG or
larger constitutes evidence in support of an upper limit of $n \gtrsim
10^8$~cm$^{-3}$ for OH maser production.  Confirmation of the
previously-noted large magnetic field in W51 Main \citep{fish06} is
therefore an important result of the present work.

We find no evidence that the magnetic field strengths in ON 1 or K3-50
are changing with time over the 4-year baseline of our observations.
Observations of Cep A over a span of more than 20 years have shown
that the magnetic field strengths in two Zeeman pairs are decreasing
with time \citep{cohen90,bartkiewicz05}.  Cohen et al.\ speculate that
the decrease in magnetic field strength is due to the expansion of the
gas around the central star.  Based on the rate of field decay, they
estimate an expansion age of 500~yr for Cep A, which agrees with the
1000~yr age estimate for the young massive stars in Cep A by
\citet{hughes85}.  Both ON 1 and K3-50 are much older SFRs, with
dynamical ages over 1000 and 5000 yr, respectively, based on the OH
maser motions obtained in this work (and also reflected in the size of
their \hii\ regions).  Thus it is not surprising that the magnetic
field strengths in our sources are not observed to vary.  As for W51
Main and South, Zeeman pairings are occasionally ambiguous due to the
clustering of maser spots.  However, the magnetic field strengths and
directions we obtain are consistent with those deduced from the first
epoch of observations 9 years previously.

\acknowledgments

The National Radio Astronomy Observatory is a facility of the National
Science Foundation operated under cooperative agreement by Associated
Universities, Inc.

{\it Facilities: VLBA}

\clearpage
\LongTables

\begin{deluxetable}{rrcrrrr}
\tablewidth{0pt}
\tabletypesize{\small}
\tablecaption{Detected Masers in ON~1\label{on1-table}}
\tablehead{
  \colhead{$\Delta$ R.A.} &
  \colhead{$\Delta$ Decl.} &
  \colhead{} &
  \colhead{$v_\mathrm{LSR}$\tablenotemark{a}} &
  \colhead{$\Delta v$\tablenotemark{a}} &
  \colhead{Brightness\tablenotemark{a}} &
  \colhead{Feature}   \\
  \colhead{(mas)} &
  \colhead{(mas)} &
  \colhead{Pol.} &
  \colhead{(km\,s$^{-1}$)} &
  \colhead{(km\,s$^{-1}$)} &
  \colhead{(Jy\,beam$^{-1}$)} &
  \colhead{Number\tablenotemark{b}}
}
\startdata
\cutinhead{1665~MHz}
  $-$965.68 &     666.00 & R & 11.89 & 0.28 &  0.72 & \tablenotemark{c} \\
  $-$965.25 &     665.92 & L & 11.88 & 0.29 &  0.66 & \nod \\
  $-$964.88 &     665.67 & L & 13.04 & \nod &  0.11 & \nod \\
  $-$884.60 &     707.53 & L & 13.75 & \nod &  0.12 & \nod \\
  $-$874.06 &     581.81 & R & 11.64 & \nod &  0.43 & \tablenotemark{c} \\
  $-$855.70 &     697.28 & L & 15.15 & 0.28 &  0.36 & \tablenotemark{c} \\
  $-$426.68 &     289.73 & R & 15.31 & 0.34 &  0.11 & \nod \\
  $-$390.64 &     209.47 & R & 14.16 & 0.44 & 26.47 &    2 \\
  $-$390.18 &     203.01 & L & 14.27 & 0.37 &  8.72 &    2 \\
  $-$382.13 &     130.66 & R & 13.57 & \nod &  0.56 &    3 \\
  $-$375.09 &     126.11 & R & 13.61 & 0.29 &  3.90 &    4 \\
  $-$331.65 &     100.54 & R & 13.32 & 0.30 &  0.42 &    6 \\
  $-$259.45 &     100.83 & R & 13.92 & \nod &  0.39 &   10 \\
  $-$258.85 &     100.60 & L & 16.73 & \nod &  0.80 &    9 \\
  $-$255.25 &      87.59 & R & 13.45 & 0.33 &  2.35 &   12 \\
  $-$255.13 &      88.20 & L & 16.47 & 0.31 &  5.37 &   11 \\
  $-$249.72 &     103.36 & R & 13.68 & 0.29 &  3.98 &   13 \\
  $-$249.24 &     102.42 & L & 16.73 & \nod &  1.35 &   14 \\
  $-$238.90 &      98.32 & R & 13.70 & 0.29 &  0.61 &   15 \\
  $-$235.05 &      96.69 & L & 16.10 & 0.38 &  4.63 & \nod \\
  $-$233.76 &      97.58 & L & 16.35 & 0.46 &  4.13 &   16 \\
  $-$212.69 &      80.57 & L & 15.86 & \nod &  0.48 &   19 \\
  $-$201.74 &      85.70 & L & 15.88 & 0.45 &  0.85 & \nod \\
  $-$128.22 &     994.63 & R &  5.95 & 0.40 &  0.37 &   22 \\
   $-$64.88 &    1024.65 & L &  3.10 & 0.29 &  0.42 &   23 \\
   $-$64.20 &    1000.55 & L &  4.95 & 0.34 &  0.71 &   24 \\
   $-$57.36 &     999.58 & L &  6.08 & 0.32 &  1.33 &   25 \\
   $-$52.13 &    1006.29 & L &  4.61 & \nod &  0.34 &   27 \\
   $-$50.68 &    1012.23 & L &  3.90 & \nod &  0.62 &   28 \\
   $-$44.39 &     995.89 & L &  4.67 & 0.43 &  0.37 &   29 \\
   $-$42.84 &    1037.19 & L &  3.49 & 0.30 & 18.03 &   30 \\
   $-$41.37 &    1052.15 & L &  5.31 & \nod &  0.19 &   31 \\
   $-$39.74 &    1059.39 & L &  5.66 & \nod &  0.11 &   33 \\
   $-$22.44 &    1109.70 & R &  2.48 & 0.32 &  0.15 &   34 \\
   $-$18.44 &    1129.78 & R &  3.50 & 0.27 &  0.49 &   35 \\
    $-$0.06 &    $-$1.22 & L & 12.85 & 0.44 &  0.27 &   36 \\
    $-$0.02 &    $-$0.02 & R & 12.82 & 0.32 & 39.59 &   36 \\
      16.49 &       5.23 & R & 14.29 & 0.28 &  0.64 & \nod \\
      18.03 &       4.86 & L & 14.31 & 0.27 &  1.15 &   37 \\
      32.64 &     532.38 & R & 13.14 & 0.40 &  0.71 &   39 \\
      38.86 &     526.62 & L & 13.36 & 0.45 &  0.39 & \nod \\
      39.19 &     527.41 & R & 13.21 & 0.48 &  0.47 &   40 \\
      40.34 &     526.99 & L & 13.88 & 0.65 &  1.08 &   41 \\
      40.40 &     527.71 & R & 13.85 & 0.40 &  0.32 & \nod \\
     152.31 &    $-$4.82 & R & 14.02 & 0.30 &  4.39 &   43 \\
     154.58 &    $-$6.23 & L & 15.05 & 0.36 &  1.43 &   44 \\
     156.30 &    $-$6.62 & L & 14.11 & 0.39 &  0.92 & \nod \\
     162.13 &    $-$7.20 & L & 14.80 & \nod &  0.40 & \nod \\
     374.87 &  $-$409.70 & L & 11.64 & \nod &  0.07 & \tablenotemark{c} \\
\cutinhead{1667~MHz}
  $-$377.57 &     125.65 & R & 14.22 & 0.28 &  1.60 &   45 \\
   $-$17.04 &    $-$8.24 & R & 13.84 & 0.28 &  1.22 &   46 \\
     160.68 &   $-$13.49 & R & 14.29 & 0.37 &  0.20 &   47 \\
     161.92 &   $-$13.37 & L & 14.64 & 0.33 &  0.56 &   48 \\
     169.53 &   $-$16.79 & R & 14.38 & 0.35 &  0.15 &   49 \\
     170.54 &   $-$16.96 & L & 14.62 & \nod &  0.34 & \nod
\enddata
\tablenotetext{a}{{}LSR velocity, line width, and brightness are fitted
  quantities when available.  When a maser has not been detected in at
  least three channels including at least one channel on either side
  of the peak, no line width is given, and the brightness and
  LSR velocity refer to the channel of brightest emission.}
\tablenotetext{b}{Corresponding maser feature number in \citet{fishvlba}.}
\tablenotetext{c}{See \S \ref{results-on1}.}
\end{deluxetable}

\vfill
\pagebreak

\begin{deluxetable}{rrcrrrr}
\tabletypesize{\small}
\tablecaption{Detected Masers in K3-50\label{k350-table}}
\tablehead{
  \colhead{$\Delta$ R.A.} &
  \colhead{$\Delta$ Decl.} &
  \colhead{} &
  \colhead{$v_\mathrm{LSR}$} &
  \colhead{$\Delta v$} &
  \colhead{Brightness} &
  \colhead{Feature} \\
  \colhead{(mas)} &
  \colhead{(mas)} &
  \colhead{Pol.} &
  \colhead{(km\,s$^{-1}$)} &
  \colhead{(km\,s$^{-1}$)} &
  \colhead{(Jy\,beam$^{-1}$)} &
  \colhead{Number}
}
\startdata
\cutinhead{1665~MHz}
   $-$113.57 &  $-$467.50 & R & $-$18.84 & 0.43 & 0.24 &    1 \\
   $-$105.41 &  $-$599.80 & L & $-$17.52 & 0.32 & 1.17 &    2 \\
    $-$89.63 &  $-$572.22 & R & $-$22.00 & \nod & 0.36 &    3 \\
    $-$72.04 &  $-$614.53 & R & $-$22.18 & \nod & 0.21 & \nod \\
    $-$50.69 &  $-$440.19 & L & $-$22.43 & 0.33 & 0.85 &    4 \\
     $-$6.12 &       1.20 & R & $-$21.32 & 0.48 & 3.25 &    5 \\
        0.00 &       0.00 & L & $-$19.75 & 0.37 & 8.16 &    6 \\
       15.41 &       3.29 & L & $-$20.39 & 0.41 & 2.70 &    7 \\
      276.37 &     155.04 & R & $-$19.51 & 0.38 & 1.03 &    8 \\
      277.41 &     152.77 & R & $-$20.92 & 0.68 & 0.28 &    9 \\
      282.44 &     149.67 & L & $-$19.31 & 0.34 & 0.77 &   10 \\
     1739.20 & $-$1921.46 & R & $-$22.18 & \nod & 0.77 &   11 \\
     1743.47 & $-$1925.42 & R & $-$23.19 & 1.07 & 0.93 &   12 \\
     1745.09 & $-$1937.14 & L & $-$22.00 & \nod & 0.65 & \nod \\
     1745.44 & $-$1925.10 & L & $-$21.43 & 1.31 & 0.79 &   13 \\
\cutinhead{1667~MHz}
        0.35 &       3.03 & R & $-$21.04 & 0.50 & 1.38 &   15 \\
        0.49 &       2.11 & L & $-$20.12 & 0.38 & 1.85 &   16 \\
       10.71 &       4.19 & L & $-$20.78 & 0.55 & 1.70 &   17
\enddata
\end{deluxetable}

\begin{deluxetable}{rrcrrrr}
\tabletypesize{\small}
\tablecaption{Detected Masers in W51~South\label{w51s-table}}
\tablehead{
  \colhead{$\Delta$ R.A.} &
  \colhead{$\Delta$ Decl.} &
  \colhead{} &
  \colhead{$v_\mathrm{LSR}$} &
  \colhead{$\Delta v$} &
  \colhead{Brightness} &
  \colhead{Feature} \\
  \colhead{(mas)} &
  \colhead{(mas)} &
  \colhead{Pol.} &
  \colhead{(km\,s$^{-1}$)} &
  \colhead{(km\,s$^{-1}$)} &
  \colhead{(Jy\,beam$^{-1}$)} &
  \colhead{Number}
}
\startdata
\cutinhead{1665~MHz}
  $-$1962.34 &     778.46 & L &   62.62 & 0.58 &   0.52 &    2 \\
  $-$1940.65 &     776.89 & L &   62.50 & 0.35 &   0.27 &    3 \\
  $-$1921.10 &     765.50 & L &   62.46 & \nod &   0.25 & \nod \\
  $-$1918.49 &     820.74 & R &   62.11 & \nod &   0.17 &    5 \\
  $-$1891.66 &     774.78 & L &   61.77 & 0.49 &   0.30 &    7 \\
  $-$1890.21 &    1065.91 & L &   60.87 & 0.41 &   2.02 &    8 \\
  $-$1886.06 &     773.72 & R &   61.70 & 0.49 &   0.22 &    9 \\
  $-$1884.54 &     773.16 & L &   62.11 & \nod &   0.27 & \nod \\
  $-$1835.12 &  $-$420.83 & R &   54.02 & \nod &   0.35 &   11 \\
  $-$1822.81 &  $-$390.73 & L &   50.68 & \nod &   0.21 &   12 \\
  $-$1821.88 &  $-$394.92 & R &   54.20 & \nod &   0.25 &   13 \\
  $-$1818.79 &    1084.92 & L &   59.82 & \nod &   0.17 & \nod \\
  $-$1717.81 &    1059.05 & R &   58.42 & \nod &   0.63 & \nod \\
  $-$1525.78 &    1162.25 & R &   57.79 & 0.38 &   0.66 &   18 \\
  $-$1524.80 &    1163.29 & L &   57.79 & 0.33 &   0.79 &   18 \\
   $-$814.99 &  $-$210.36 & R &   59.61 & 0.43 &   3.15 & \nod \\
   $-$770.25 &  $-$222.97 & L &   59.75 & 0.47 &   0.29 & \nod \\
   $-$769.59 &  $-$221.27 & R &   59.74 & 0.46 &   0.64 & \nod \\
   $-$624.12 &    1049.92 & R &   56.13 & \nod &   0.28 &   19 \\
   $-$619.48 &    1081.61 & R &   56.65 & 0.53 &   0.47 &   20 \\
   $-$578.14 &    1108.77 & L &   57.71 & \nod &   0.20 & \nod \\
   $-$565.42 &    1199.88 & R &   55.92 & 0.43 &   1.04 &   21 \\
   $-$562.75 &    1201.05 & L &   59.91 & 0.65 &   0.23 & \nod \\
   $-$548.58 &    1231.79 & R &   56.49 & 0.28 &   0.41 & \nod \\
   $-$495.74 & $-$1235.36 & R &   62.78 & 0.36 &   0.48 & \nod \\
   $-$432.63 & $-$1172.43 & L &   61.47 & 0.85 &   0.33 & \nod \\
   $-$421.34 & $-$1151.00 & R &   61.23 & \nod &   0.32 &   24 \\
   $-$420.06 & $-$1151.32 & L &   61.21 & 0.30 &   0.44 &   24 \\
   $-$415.15 &  $-$885.46 & R &   61.58 & \nod &   0.15 & \nod \\
   $-$392.77 &  $-$995.12 & L &   63.16 & \nod &   0.53 & \nod \\
   $-$359.43 & $-$1851.60 & L &   57.13 & 0.32 &   1.59 &   26 \\
   $-$342.14 & $-$1829.88 & L &   56.75 & 0.56 &   1.17 &   28 \\
   $-$263.84 & $-$1176.27 & R &   62.15 & 0.32 &   1.88 &   29 \\
   $-$261.66 & $-$1174.74 & L &   59.28 & 0.30 &   0.46 & \nod \\
   $-$230.97 & $-$1781.71 & R &   61.58 & \nod &   0.18 &   30 \\
   $-$221.88 & $-$1790.79 & R &   58.59 & \nod &   0.88 & \nod \\
   $-$215.40 &  $-$799.30 & R &   60.35 & \nod &   0.15 & \nod \\
   $-$192.30 &      31.61 & L &   61.05 & \nod &   0.17 & \nod \\
   $-$136.49 &  $-$809.77 & L &   61.72 & 0.39 &   0.85 & \nod \\
   $-$136.34 &  $-$808.96 & R &   61.69 & 0.36 &   0.93 & \nod \\
   $-$128.68 &  $-$817.75 & L &   62.76 & 0.79 &   0.11 & \nod \\
   $-$126.58 &  $-$817.46 & R &   62.89 & 0.57 &   0.25 &   33 \\
   $-$124.10 &  $-$816.70 & L &   61.61 & 0.41 &   1.79 &   34 \\
   $-$123.69 &  $-$815.99 & R &   61.58 & 0.43 &   1.24 &   35 \\
    $-$69.84 &   $-$54.33 & R &   60.11 & 0.45 &   0.71 &   36 \\
    $-$69.48 &   $-$56.34 & L &   60.07 & 0.41 &   0.88 &   36 \\
    $-$40.19 &   $-$70.98 & R &   59.82 & \nod &   0.26 & \nod \\
    $-$32.75 & $-$1255.39 & L &   59.12 & \nod &   0.26 & \nod \\
    $-$12.66 &      33.67 & L &   57.71 & \nod &   0.48 & \nod \\
    $-$11.80 &      30.08 & R &   57.53 & 0.53 &   0.66 & \nod \\
    $-$10.76 &   $-$12.29 & L &   59.30 & \nod &   0.51 & \nod \\
     $-$6.68 &  $-$103.27 & L &   60.63 & 0.63 &   0.24 & \nod \\
     $-$6.49 &  $-$100.55 & R &   60.74 & 0.54 &   0.19 & \nod \\
     $-$2.36 &   $-$45.31 & L &   60.56 & 0.56 &   0.23 &   38 \\
        1.64 &    $-$9.35 & L &   58.94 & 0.50 &   8.12 & \nod \\
        4.22 &       1.35 & L &   58.47 & 0.40 &  20.77 &   40 \\
        4.37 &       1.70 & R &   58.33 & 0.53 &  87.63 &   39 \\
        6.83 &  $-$857.03 & R &   60.39 & 0.55 &   0.22 & \nod \\
        7.71 &   $-$13.36 & L &   58.40 & 0.49 &  10.57 & \nod \\
        8.54 &  $-$857.50 & L &   60.33 & 0.48 &   0.23 & \nod \\
       22.69 &      27.99 & L &   57.54 & \nod &   0.38 & \nod \\
       26.15 &      11.42 & R &   57.43 & 0.43 &   1.91 & \nod \\
       28.50 &    $-$1.21 & R &   59.72 & 0.56 &   1.87 & \nod \\
       29.26 &       5.01 & L &   58.37 & 0.37 &   4.64 & \nod \\
       29.37 &   $-$23.97 & R &   58.81 & 0.40 &   4.67 & \nod \\
       29.46 &   $-$23.42 & L &   58.92 & 0.58 &  19.20 &   41 \\
       29.65 &   $-$21.53 & L &   59.12 & \nod &   8.68 & \nod \\
       30.53 &  $-$150.64 & L &   61.23 & \nod &   0.15 & \nod \\
       36.24 &   $-$66.80 & R &   60.77 & 0.38 &   1.72 & \nod \\
       39.00 &   $-$58.09 & R &   58.79 & 0.39 &  18.21 &   44 \\
       42.96 &   $-$56.92 & L &   58.93 & 0.36 &   3.98 & \nod \\
       43.40 &   $-$64.55 & L &   62.45 & 0.37 &   0.56 & \nod \\
       43.96 &   $-$64.28 & R &   62.48 & 0.37 &   0.21 & \nod \\
       51.90 &   $-$14.37 & R &   60.35 & \nod &   0.17 & \nod \\
       55.38 &    $-$0.95 & L &   60.02 & 0.45 &   0.72 & \nod \\
       58.70 &   $-$55.33 & L &   59.82 & \nod &   0.30 & \nod \\
       61.30 &   $-$66.28 & R &   61.23 & \nod &   0.28 & \nod \\
       61.74 &   $-$40.83 & L &   59.50 & 0.47 &   0.97 & \nod \\
       86.16 &       8.48 & L &   60.07 & 0.42 &   0.46 & \nod \\
       94.45 &   $-$27.96 & L &   57.54 & 0.32 &   0.65 &   45 \\
      109.52 &      11.38 & L &   59.80 & 0.93 &   0.36 & \nod \\
      113.85 &      14.92 & R &   59.26 & 0.64 &   0.28 & \nod \\
      126.04 &    $-$0.85 & L &   58.73 & 0.44 &   3.95 & \nod \\
      140.12 &     237.55 & L &   61.70 & 0.31 &   2.65 &   47 \\
      140.63 &     281.62 & L &   61.49 & 0.39 &   1.76 &   48 \\
      148.16 &      48.45 & L &   57.76 & 0.61 &   1.52 &   49 \\
      149.25 &       5.37 & R &   60.88 & \nod &   0.46 & \nod \\
      263.20 &  $-$170.93 & R &   60.88 & \nod &   0.26 & \nod \\
      264.15 &  $-$172.06 & L &   60.88 & 0.33 &   0.51 & \nod \\
      300.88 &  $-$158.62 & R &   57.73 & 0.46 &   1.28 & \nod \\
      320.58 &     744.94 & R &   64.86 & 0.33 &   1.07 &   51 \\
      321.54 &     743.94 & L &   64.86 & 0.36 &   0.77 &   51 \\ 
      339.04 &     201.10 & L &   59.82 & \nod &   0.17 & \nod \\
      347.71 &  $-$225.22 & L &   59.97 & 0.31 &   0.51 & \nod \\
      347.91 &  $-$225.07 & R &   60.00 & \nod &   0.58 & \nod \\
\cutinhead{1667~MHz}
   $-$815.34 &  $-$213.27 & R &   59.65 & \nod &   0.23 & \nod \\
   $-$785.81 &  $-$228.73 & R &   59.63 & 0.58 &   1.26 & \nod \\
   $-$785.54 &  $-$229.00 & L &   59.59 & 0.54 &   0.69 & \nod \\
   $-$769.05 &  $-$225.89 & L &   59.88 & 0.54 &   0.53 & \nod \\
   $-$768.42 &  $-$225.04 & R &   59.84 & 0.51 &   1.03 & \nod \\
   $-$419.86 & $-$1155.51 & L &   61.76 & \nod &   0.27 & \nod \\
   $-$110.62 &  $-$822.13 & R &   61.35 & 0.32 &   0.77 & \nod \\
   $-$110.15 &  $-$823.11 & L &   61.37 & 0.33 &   0.98 &   65 \\
   $-$109.60 &  $-$823.86 & R &   62.11 & \nod &   0.82 &   66 \\
       70.35 &   $-$51.67 & L &   61.96 & 0.41 &   2.52 &   69 \\
       74.82 &   $-$53.80 & R &   60.68 & 0.41 &   1.13 &   71 \\
      133.05 &     234.23 & R &   60.18 & \nod &   0.16 &   73 \\
      139.20 &     234.27 & L &   61.21 & 0.28 &   0.53 &   74 \\
      155.91 &    $-$8.29 & R &   61.62 & 0.69 &   0.16 &   75 \\
      170.67 &       1.26 & L &   62.51 & 0.48 &   0.23 & \nod \\
      212.82 &      15.67 & R &   62.59 & 0.46 &   0.45 &   79 \\
      221.73 &     640.71 & L &   60.00 & 0.29 &   0.28 &   80 \\
      352.19 &     206.06 & R &   59.92 & 0.61 &   0.22 &   81 \\ 
      360.44 &     206.65 & L &   59.93 & 0.67 &   0.19 &   81 \\
      431.07 &  $-$375.19 & R &   61.95 & 0.39 &   0.40 &   82 \\
      449.12 &  $-$360.57 & L &   61.96 & 0.41 &   2.02 &   83 \\
      450.65 &  $-$360.89 & R &   62.03 & 0.37 &   3.73 &   84 \\
      465.85 &  $-$340.96 & R &   61.96 & 0.37 &   2.18 &   87 \\
      466.98 &  $-$342.06 & L &   61.92 & 0.38 &   1.40 &   86 \\
      542.60 &  $-$368.86 & R &   61.40 & \nod &   0.14 & \nod \\
      543.81 &  $-$374.97 & R &   60.92 & 0.48 &   0.21 &   90 \\
      545.72 &  $-$373.57 & L &   60.87 & 0.51 &   0.30 &   91 \\
      557.62 &  $-$331.19 & L &   60.88 & \nod &   0.17 &   92 \\
      575.59 &  $-$201.30 & R &   61.89 & 0.35 &   1.49 &   96 \\
      575.95 &  $-$202.37 & L &   61.84 & 0.49 &   0.65 &   96 \\
\cutinhead{1720~MHz}
  $-$2071.85 &  $-$363.09 & R &   55.43 & 0.41 &   2.19 & \nod \\
  $-$2070.55 &  $-$364.96 & L &   54.60 & 0.37 &   1.07 & \nod \\
  $-$2046.53 &  $-$392.14 & R &   56.64 & 0.49 &   0.38 & \nod \\
  $-$1862.64 &  $-$419.83 & R &   53.56 & 0.52 &   1.56 & \nod \\
  $-$1857.76 &  $-$420.17 & L &   52.65 & 0.36 &   1.22 & \nod \\
  $-$1839.64 &  $-$420.68 & R &   53.36 & \nod &   0.67 & \nod 
\enddata
\end{deluxetable}

\begin{deluxetable}{rrcrrrr}
\tabletypesize{\small}
\tablecaption{Detected Masers in W51~Main\label{w51m-table}}
\tablehead{
  \colhead{$\Delta$ R.A.\tablenotemark{a}} &
  \colhead{$\Delta$ Decl.\tablenotemark{a}} &
  \colhead{} &
  \colhead{$v_\mathrm{LSR}$} &
  \colhead{$\Delta v$} &
  \colhead{Brightness} &
  \colhead{Feature} \\
  \colhead{(mas)} &
  \colhead{(mas)} &
  \colhead{Pol.} &
  \colhead{(km\,s$^{-1}$)} &
  \colhead{(km\,s$^{-1}$)} &
  \colhead{(Jy\,beam$^{-1}$)} &
  \colhead{Number}
}
\startdata
\cutinhead{1665~MHz}
  $-$1182.34 &    5755.96 & L &   49.31 & 0.45 &   0.50 &    1 \\
  $-$1170.13 &    5786.08 & L &   48.57 & \nod &   0.18 &    2 \\
  $-$1147.87 &    5702.08 & L &   50.33 & \nod &   0.16 &    3 \\
  $-$1113.49 &    6124.29 & R &   50.52 & 0.42 &   1.04 &    4 \\
  $-$1085.33 &    5219.76 & L &   54.69 & 0.54 &   0.55 &    6 \\
  $-$1085.27 &    5221.60 & R &   57.01 & 0.44 &   0.46 &    7 \\
  $-$1067.94 &    5249.91 & R &   60.10 & 0.74 &   0.77 &    8 \\
  $-$1065.40 &    5250.17 & L &   57.75 & 0.75 &   0.56 &    9 \\
   $-$997.36 &    5353.95 & R &   61.93 & \nod &   0.20 & \nod \\
   $-$855.25 &    5319.67 & L &   52.27 & \nod &   0.21 & \nod \\
   $-$854.02 &    5624.93 & R &   57.22 & 0.45 &   2.61 &   15 \\
   $-$849.04 &    5697.37 & L &   49.42 & 0.33 &   1.14 &   18 \\
   $-$834.81 &    5650.98 & R &   57.98 & 0.74 &   1.49 &   19 \\
   $-$825.21 &    5645.36 & R &   55.89 & 0.21 &   9.96 &   20 \\
   $-$823.99 &    5643.76 & L &   55.96 & \nod &   0.24 & \nod \\
   $-$815.12 &    5659.66 & R &   59.12 & \nod &   0.89 & \nod \\
   $-$783.24 &    5674.08 & R &   59.39 & 0.47 &  16.29 & \nod \\
   $-$772.70 &    5638.75 & R &   58.80 & 0.66 &  31.89 &   21 \\
   $-$707.91 &    5667.86 & R &   57.73 & 0.50 &   4.62 &   24 \\
   $-$700.66 &    5649.71 & R &   57.31 & 0.72 &   2.02 &   25 \\
   $-$697.88 &    5648.14 & R &   56.27 & 0.65 &   0.82 & \nod \\
   $-$694.30 &    6156.55 & R &   64.22 & 0.49 &   0.83 & \nod \\
   $-$693.91 &    6156.95 & L &   51.91 & \nod &   0.17 &   26 \\
   $-$690.98 &    6187.16 & R &   67.60 & 0.41 &   1.05 &   29 \\
   $-$689.63 &    6188.76 & L &   55.61 & \nod &   0.18 &   28 \\
   $-$444.88 &    6611.47 & L &   68.90 & 1.14 &   0.22 &   31 \\
   $-$444.50 &    6609.96 & R &   71.43 & 0.49 &   0.23 &   32 \\
   $-$301.22 &    6282.29 & R &   64.20 & 0.46 &   0.31 &   34 \\
   $-$274.83 &    6558.57 & R &   67.21 & 0.31 &   0.27 &   35 \\
   $-$232.59 &    6512.03 & L &   67.73 & 0.36 &   0.65 &   36 \\
   $-$232.52 &    6511.52 & R &   70.20 & \nod &   0.26 &   37 \\
   $-$229.26 &    6524.31 & L &   67.03 & \nod &   0.14 &   39 \\
   $-$223.82 &    6712.15 & R &   69.73 & 0.42 &   0.63 &   38 \\
   $-$180.44 &    5925.77 & L &   54.94 & 0.36 &   1.52 &   40 \\
   $-$127.57 &    6581.97 & R &   65.50 & 0.63 &   0.14 &   41 \\
   $-$109.08 &    5896.89 & L &   52.09 & \nod &   0.30 & \nod \\
   $-$106.25 &    6592.58 & R &   64.59 & 0.46 &   1.97 &   42 \\
   $-$105.86 &    6592.52 & L &   64.61 & 0.43 &   0.89 &   42 \\
   $-$100.47 &    6603.44 & L &   64.22 & \nod &   0.68 & \nod \\
    $-$95.12 &    5870.46 & L &   54.23 & 0.29 &   0.87 &   43 \\
    $-$91.34 &    5856.99 & R &   60.18 & \nod &   0.53 & \nod \\
    $-$31.04 &    5478.85 & R &   57.71 & \nod &   0.37 &   46 \\
    $-$28.64 &    5478.92 & L &   51.85 & 0.45 &   0.48 &   45 \\
        6.95 &    5839.69 & L &   52.99 & 0.37 &   0.30 & \nod \\
       23.54 &    6332.59 & R &   61.64 & 0.52 &   0.23 & \nod \\
       42.88 &    5824.34 & L &   52.44 & \nod &   0.24 &   47 \\
      106.83 &    5714.90 & L &   46.70 & 0.41 &   0.94 &   50 \\
      152.33 &    5788.58 & L &   46.66 & 0.35 &   0.25 &   52 \\
      400.73 &    5729.79 & L &   47.92 & 0.47 &   0.50 &   53 \\
\cutinhead{1667~MHz}
  $-$1091.99 &    5260.30 & L &   60.53 & \nod &   0.23 &   54 \\
  $-$1081.93 &    5302.03 & L &   59.54 & 0.64 &   0.46 &   57 \\
  $-$1080.66 &    5297.19 & R &   61.76 & 0.98 &   0.41 & \nod \\
  $-$1079.67 &    5295.13 & L &   60.35 & 0.74 &   0.29 & \nod \\
  $-$1077.29 &    5292.93 & R &   63.43 & 0.81 &   0.52 &   58 \\
  $-$1076.96 &    5292.84 & L &   61.91 & 0.74 &   0.61 &   59 \\
  $-$1008.01 &    5335.45 & R &   62.27 & 0.57 &   0.61 &   65 \\
  $-$1004.65 &    5334.97 & L &   60.72 & 0.49 &   0.59 & \nod \\
   $-$886.62 &    6633.63 & L &   70.08 & 0.61 &   0.24 &   67 \\
   $-$792.72 &    5694.23 & R &   60.00 & \nod &   0.17 & \nod \\
   $-$706.73 &    6069.73 & L &   54.16 & 0.32 &  10.04 &   71 \\
   $-$693.33 &    5651.72 & R &   56.03 & 0.49 &   1.46 &   72 \\
   $-$692.61 &    5650.86 & L &   55.96 & \nod &   0.19 &   72 \\
   $-$666.96 &    6579.55 & R &   69.66 & \nod &   0.19 &   73 \\
   $-$653.01 &    6041.47 & L &   55.25 & 0.33 &   0.19 & \nod \\
   $-$300.44 &    6275.75 & R &   63.16 & \nod &   0.17 & \nod \\
   $-$299.89 &    6270.42 & R &   63.94 & 0.44 &   0.32 &   77 \\
   $-$273.97 &    6555.48 & R &   66.65 & 0.40 &   0.44 &   78 \\
   $-$232.14 &    6507.51 & R &   69.73 & 0.34 &   2.64 &   80 \\
   $-$231.80 &    6507.07 & L &   68.32 & 0.35 &   1.04 &   79 \\
   $-$231.59 &    6506.25 & L &   69.72 & 0.31 &   0.71 & \nod \\
   $-$220.38 &    6707.61 & R &   68.79 & 0.63 &   1.53 &   82 \\
   $-$180.82 &    5922.92 & L &   56.76 & 0.42 &   0.81 &   84 \\
   $-$108.89 &    5893.63 & L &   53.15 & \nod &   0.17 & \nod \\
   $-$108.31 &    5894.74 & R &   57.30 & 0.57 &   0.35 & \nod \\
   $-$104.91 &    6588.53 & R &   64.92 & 0.56 &   1.08 &   86 \\
      102.45 &    5711.74 & R &   49.35 & 0.53 &   0.20 & \nod \\
      104.61 &    5711.23 & L &   47.36 & 0.49 &   0.54 &   88 \\
      130.08 &    5768.24 & L &   46.99 & 0.55 &   1.55 &   90 \\
      130.29 &    5766.44 & R &   48.83 & 0.55 &   0.67 &   91 \\
      152.60 &    5789.24 & R &   49.11 & \nod &   0.20 &   92 \\
      156.11 &    5788.40 & L &   47.18 & \nod &   0.24 &   93 \\
      180.66 &    5813.92 & R &   48.59 & \nod &   0.17 &   94 \\
\cutinhead{1720~MHz}
  $-$1220.45 &    5336.56 & R &   55.08 & 0.38 &   3.78 & \nod \\
  $-$1218.96 &    5334.32 & L &   54.31 & 0.41 &   1.21 & \nod \\
  $-$1105.77 &    5502.33 & R &   56.72 & 1.12 &   1.27 & \nod \\
  $-$1103.96 &    5497.80 & L &   55.85 & 0.69 &   0.76 & \nod \\
  $-$1066.39 &    5441.33 & L &   58.13 & \nod &   0.26 & \nod \\
   $-$978.82 &    5362.48 & R &   60.60 & 0.61 &   0.38 & \nod \\
   $-$953.47 &    5189.18 & R &   55.39 & 0.47 &   9.62 & \nod \\
   $-$951.36 &    5188.07 & L &   54.60 & 0.50 &   2.61 & \nod \\
   $-$950.25 &    5530.69 & L &   58.71 & 0.52 &   0.91 & \nod \\
   $-$941.99 &    5539.44 & L &   59.53 & 0.52 &   4.62 & \nod \\
   $-$941.96 &    5543.05 & R &   60.43 & 0.57 &  12.98 & \nod \\
   $-$936.60 &    5688.29 & R &   53.64 & 0.79 &   2.53 & \nod \\
   $-$932.87 &    5686.24 & L &   52.45 & 0.47 &   1.68 & \nod \\
   $-$916.17 &    5665.59 & R &   56.12 & 0.40 &  31.36 & \nod \\
   $-$914.04 &    5663.76 & L &   55.16 & 0.52 &   7.45 & \nod \\
   $-$910.29 &    5259.37 & R &   56.51 & 0.45 &   3.88 & \nod \\
   $-$907.75 &    5257.91 & L &   55.79 & 0.49 &   3.22 & \nod \\
   $-$899.04 &    5582.36 & R &   59.58 & 0.60 & 131.13 & \nod \\
   $-$897.48 &    5580.55 & L &   58.67 & 0.51 &  33.47 & \nod \\
   $-$886.67 &    5234.46 & L &   55.03 & 0.32 &   0.96 & \nod \\
   $-$886.18 &    5234.40 & R &   55.75 & 0.30 &   1.35 & \nod \\
   $-$867.16 &    5289.35 & R &   55.86 & 0.63 &   0.77 & \nod \\
   $-$863.24 &    5290.85 & L &   54.87 & 0.63 &   0.64 & \nod \\
   $-$844.12 &    5618.13 & L &   56.86 & 0.76 &  10.93 & \nod \\
   $-$838.18 &    5625.31 & R &   58.04 & 0.99 &  27.84 & \nod \\
   $-$832.98 &    5591.20 & R &   61.00 & 0.48 &   1.34 & \nod \\
   $-$831.51 &    5590.62 & L &   60.02 & 0.42 &   0.79 & \nod 
\enddata
\tablenotetext{a}{Offsets are computed from the origin in W51~South.}
\end{deluxetable}


\begin{thebibliography}{}
{
\bibitem[Argon et al.(2000)]{argon00} Argon, A.~L., Reid, M.~J., \&
  Menten, K.~M.\ 2000, \apjs, 129, 159

\bibitem[Araya et al.(2002)]{araya02} Araya, E., Hofner, P.,
  Churchwell, E., \& Kurtz, S.\ 2002, \apjs, 138, 63

\bibitem[Baldwin et al.(1973)]{baldwin73} Baldwin, J.~E.,
  Harris, C.~S., \& Ryle, M.\ 1973, \nat, 241, 38

\bibitem[Bartkiewicz et al.(2005)]{bartkiewicz05} Bartkiewicz, A.,
  Szymczak, M., Cohen, R.~J., \& Richards, A.~M.~S.\ 2005, \mnras,
  361, 623

\bibitem[Baudry et al.(1997)]{baudry97} Baudry, A., Desmurs, J.~F.,
  Wilson, T.~L., \& Cohen, R.~J.\ 1997, \aap, 325, 255

\bibitem[Beckwith et al.(1990)]{beckwith90} Beckwith, S.~V.~W.,
  Sargent, A.~I., Chini, R.~S., \& G\"{u}sten, R.\ 1990, \aj, 99, 924

\bibitem[Benson et al.(1984)]{benson84} Benson, J.~M., Mutel, R.~L.,
  \& Gaume, R.~A.\ 1984, \aj, 89, 1391

\bibitem[Berulis \& Ershov(1983)]{berulis83}  Berulis, I.~I., \& 
  Ershov, A.~A.\ 1983, Soviet Astronomy Letters, 9, 341 

\bibitem[Bloemhof et al.(1992)]{bloemhof92} Bloemhof, E.~E., Reid, 
  M.~J., \& Moran, J.~M.\ 1992, \apj, 397, 500 

\bibitem[Bloemhof et al.(1996)]{bloemhof96} Bloemhof, E.~E., 
  Moran, J.~M., \& Reid, M.~J.\ 1996, \apjl, 467, L117 

\bibitem[Brown \& Cragg(1991)]{brown91} Brown, R.~D., \& Cragg, D.~M.\
  1991, \apj, 378, 445

\bibitem[Carral et al.(1997)]{carral97} Carral, P., Kurtz, S.~E.,
  Rodr\'{\i}guez, L.~F., de Pree, C., \& Hofner, P.\ 1997, \apjl, 486,
  L103

\bibitem[Churchwell et al.(1990)]{churchwell90} Churchwell, E.,
  Walmsley, C.~M., \& Cesaroni, R.\ 1990, \aaps, 83, 119

\bibitem[Cohen et al.(1990)]{cohen90} Cohen, R.~J., Brebner, G.~C., \&
  Potter, M.~M.\ 1990, \mnras, 246, 3P
 
\bibitem[Cragg et al.(2002)]{cragg02} Cragg, D.~M., Sobolev, A.~M., \&
  Godfrey, P.~D.\ 2002, \mnras, 331, 521

\bibitem[Crutcher(1999)]{crutcher99} Crutcher, R.~M.\ 1999, \apj, 
520, 706 

\bibitem[de Pree et al.(1994)]{depree94} de Pree, C.~G., Goss, W.~M.,
  Palmer, P., \& Rubin, R.~H.\ 1994, \apj, 428, 670

\bibitem[de Pree et al.(1995)]{depree95} de Pree, C.~G., 
  Rodr\'{\i}guez, L.~F., \& Goss, W.~M.\ 1995, \rmxaa, 31, 39

\bibitem[Desmurs \& Baudry(1998)]{desmurs98} Desmurs, J.~F., \&
  Baudry, A.\ 1998, \aap, 340, 521

\bibitem[Eisner et al.(2002)]{eisner02} Eisner, J.~A., Greenhill,
  L.~J., Herrnstein, J.~R., Moran, J.~M., \& Menten, K.~M.\ 2002,
  \apj, 569, 334

\bibitem[Elitzur \& de Jong(1978)]{elitzur78} Elitzur, M., \& de Jong,
  T.\ 1978, \aap, 67, 323

\bibitem[Fiebig \& G\"{u}sten(1989)]{fiebig89} Fiebig, D., \&
  G\"{u}sten, R.\ 1989, \aap, 214, 333

\bibitem[Fish et al.(2006a)]{bf086} Fish, V.~L.,
  Brisken, W.~F., \& Sjouwerman, L.~O.\ 2006a, \apj, 647, 418

\bibitem[Fish \& Reid(2006)]{fish06} Fish, V.~L., \& Reid, M.~J.\
  2006, \apjs, 164, 99

\bibitem[Fish \& Reid(2007)]{fishw75n} Fish, V.~L., \& Reid, M.~J.\
  2007, \apj, 656, 952

\bibitem[Fish et al.(2003)]{fish03} Fish, V.~L., Reid, M.~J., Argon,
  A.~L., \& Menten, K.~M.\ 2003, \apj, 596, 328

\bibitem[Fish et al.(2005a)]{fishvlba} Fish, V.~L., Reid, M.~J.,
  Argon, A.~L., \& Zheng, X.-W.\ 2005a, \apjs, 160, 220

\bibitem[Fish et al.(2005b)]{fish05} Fish, V.~L., Reid, M.~J.,
  \& Menten, K.~M.\ 2005b, \apj, 623, 269

\bibitem[Fish et al.(2006b)]{fish07} Fish, V.~L., Reid, M.~J., Menten,
  K.~M., \& Pillai, T.\ 2006b, \aap, 458, 485

\bibitem[Fish \& Sjouwerman(2007)]{fishef015} Fish, V.~L., \&
  Sjouwerman, L.~O.\ 2007, \apj, in press, arXiv: 0706.3314

\bibitem[Gaume et al.(1993)]{gaume93} Gaume, R.~A.,
  Johnston, K.~J., \& Wilson, T.~L.\ 1993, \apj, 417, 645

\bibitem[Gaume \& Mutel(1987)]{gaume87} Gaume, R.~A., \& Mutel, R.~L.\
  1987, \apjs, 65, 193

\bibitem[Genzel \& Downes(1977)]{genzel77} Genzel, R., \& Downes, D.\
  1977, \aaps, 30, 145

\bibitem[Genzel et al.(1981a)]{genzel81a} Genzel, R., Reid, M.~J., 
  Moran, J.~M., \& Downes, D.\ 1981a, \apj, 244, 884 

\bibitem[Genzel et al.(1981b)]{genzel81b} Genzel, R., et al.\ 
  1981b, \apj, 247, 1039 

\bibitem[Gwinn(1994)]{gwinn94} Gwinn, C.~R.\ 1994, \apj, 429, 241

\bibitem[Habing \& Israel(1979)]{habing79} Habing, H.~J., \& Israel,
  F.~P.\ 1979, \araa, 17, 345

\bibitem[Harju et al.(1993)]{harju93} Harju, J., Walmsley, C.~M., \&
  Wouterloot, J.~G.~A.\ 1993, \aaps, 98, 51

\bibitem[Harris(1975)]{harris75} Harris, S.\ 1975, \mnras, 170, 139

\bibitem[Hildebrand(1983)]{hildebrand83} Hildebrand, R.~H.\ 1983,
  \qjras, 24, 267

\bibitem[Ho et al.(1983)]{ho83} Ho, P.~T.~P., Das, A., \&
  Genzel, R.\ 1983, \apj, 266, 596

\bibitem[Ho \& Young(1996)]{ho96} Ho, P.~T.~P., \& Young, L.~M.\ 1996,
  \apj, 472, 742

\bibitem[Hofmann et al.(2004)]{hofmann04} Hofmann, K.-H., Balega,
  Y.~Y., Preibisch, T., \& Weigelt, G.\ 2004, \aap, 417, 981

\bibitem[Hofner \& Churchwell(1996)]{hofner96} Hofner, P., \&
  Churchwell, E.\ 1996, \aaps, 120, 283

\bibitem[Hughes(1985)]{hughes85} Hughes, V.~A.\ 1985, \apj, 298, 830

\bibitem[H\"{u}ttemeister et al.(1993)]{huttemeister93}
  H\"{u}ttemeister, S. Wilson, T.~L., Henkel, C., \& Mauersberger, R.\
  1993, \aap, 276, 445

\bibitem[Imai et al.(2002)]{imai02} Imai, H., Watanabe, T., Omodaka,
  T., Nishio, M., Kameya, O., Miyaji, T., \& Nakajima, J.\ 2002,
  \pasj, 54, 741

\bibitem[Israel \& Wootten(1983)]{israel83} Israel, F.~P., \& Wootten,
  H.~A.\ 1983, \apj, 266, 580

\bibitem[Kawamura \& Masson(1998)]{kawamura98} Kawamura, J.~H., \&
  Masson, C.~R.\ 1998, \apj, 509, 720

\bibitem[Keto et al.(1995)]{keto95} Keto, E.~R., Welch, W.~J., Reid,
  M.~J., \& Ho, P.~T.~P.\ 1995, \apj, 444, 765

\bibitem[Kurtz \& Franco(2002)]{kurtz02} Kurtz, S., \& Franco, J.\
  2002, \rmxaa~Ser.~Conf., 12, 16

\bibitem[Lai et al.(2001)]{lai01} Lai, S.-P., Crutcher, R.~M., Girart,
  J.~M., \& Rao, R.\ 2001, \apj, 561, 864

\bibitem[Madden et al.(1986)]{madden86} Madden, S.~C., Irvine, W.~M.,
  Matthews, H.~E., Brown, R.~D., \& Godfrey, P.~D.\ 1986, \apj, 300,
  L79

\bibitem[Mauersberger et al.(1988)]{mauersberger88} Mauersberger, R.,
  Henkel, C., \& Wilson, T.~L.\ 1988, \aap, 205, 235

\bibitem[Mehringer(1994)]{mehringer94} Mehringer, D.~M.\ 1994, \apjs,
  91, 713

\bibitem[Migenes(1989)]{migenes89} Migenes, V.\ 1989, Ph.D.~Thesis

\bibitem[Migenes et al.(1992)]{migenes92} Migenes, V., Cohen, 
  R.~J., \& Brebner, G.~C.\ 1992, \mnras, 254, 501 

\bibitem[Minier et al.(2001)]{minier01} Minier, V., Booth, R.~S.,
  Ellingsen, S.~P., Conway, J.~E., \& Pestalozzi, M.~R.\ 2001,
  Proceedings of the 5th European VLBI Network Symposium, Chalmers
  University of Technology, Gothenburg, Sweden, eds.~J.E.~Conway,
  A.G.~Polatidis, R.S.~Booth, Y.~Pihlstr\"{o}m, 178

\bibitem[Moscadelli et al.(1999)]{moscadelli99} Moscadelli, L.,
  Menten, K.~M., Walmsley, C.~M., \& Reid, M.~J.\ 1999, \apj, 519, 244

\bibitem[Moscadelli et al.(2002)]{moscadelli02} Moscadelli, L.,
  Menten, K.~M., Walmsley, C.~M., \& Reid, M.~J.\ 2002, \apj, 564, 813

\bibitem[Nammahachak et al.(2006)]{nammahachak06} Nammahachak, S.,
  Asanok, K., Hutawarakorn Kramer, B., Cohen, R.~J., Muanwong, O., \&
  Gasiprong, N.\ 2006, \mnras, 371, 619

\bibitem[Okamoto et al.(2003)]{okamoto03} Okamoto, Y.~K., Kataza, H.,
  Yamashita, T., Miyata, T., Sako, S., Takubo, S., Honda, M., \&
  Onaka, T.\ 2003, \apj, 584, 368

\bibitem[Pavlakis \& Kylafis(1996)]{pavlakis96} Pavlakis, K.~G., \&
  Kylafis, N.~D.\ 1996, \apj, 467, 309

\bibitem[Phillips \& van Langevelde(2005)]{phillips05} Phillips,
  C.~J., \& van Langevelde, H.~J.\ 2005, \apss, 295, 225

\bibitem[Pratap et al.(1991)]{pratap91} Pratap, P., Menten, K.~M.,
  Reid, M.~J., Moran, J.~M., \& Walmsley, C.~M.\ 1991, \apjl, 373, L13

\bibitem[Reid et al.(1980)]{reid80} Reid, M.~J., Haschick, A.~D.,
  Burke, B.~F., Moran, J.~M., Johnston, K.~J., \& Swenson, G.~W., Jr.\
  1980, \apj, 239, 89

\bibitem[Rudolph et al.(1990)]{rudolph90} Rudolph, A., Welch, W.~J.,
  Palmer, P., \& Dubrulle, B.\ 1990, \apj, 363, 528

\bibitem[Scott(1978)]{scott78} Scott, P.~F.\ 1978, \mnras, 183, 435

\bibitem[Shepherd et al.(1997)]{shepherd97} Shepherd,
  D.~S., Churchwell, E., \& Wilner, D.~J.\ 1997, \apj, 482, 355

\bibitem[Shu(1992)]{shu92} Shu, F.~H.\ 1992, Physics of Astrophysics,
  Vol.~II (Sausalito, CA: University Science Books)

\bibitem[Sollins et al.(2004)]{sollins04} Sollins, P.~K.,
  Zhang, Q., \& Ho, P.~T.~P.\ 2004, \apj, 606, 943

\bibitem[Vogel et al.(1987)]{vogel87} Vogel, S.~N., Genzel, R., \&
  Palmer, P.\ 1987, \apj, 316, 243

\bibitem[Ward-Thompson et al.(1995)]{ward-thompson95} Ward-Thompson,
  D., Chini, R., Kr\"{u}gel, E., Andr\'{e}, P., \& Bontemps, S.\ 1995,
  \mnras, 274, 1219

\bibitem[Welch \& Marr(1987)]{welch87} Welch, W.~J., \& Marr, J.\
  1987, \apjl, 317, L21

\bibitem[Wilson \& Henkel(1988)]{wilson88} Wilson, T.~L., \& Henkel,
  C.\ 1988, \aap, 206, L26

\bibitem[Wright et al.(2004)]{wright04a} Wright, M.~M., Gray, M.~D.,
  \& Diamond, P.~J.\ 2004, \mnras, 350, 1253

\bibitem[Zhang \& Ho(1997)]{zhang97} Zhang, Q., \& Ho, P.~T.~P.\ 1997,
  \apj, 488, 241

\bibitem[Zhang et al.(1998)]{zhang98} Zhang, Q., Ho, P.~T.~P.,
  \& Ohashi, N.\ 1998, \apj, 494, 636

\bibitem[Zheng et al.(1985)]{zheng85} Zheng, X.~W., Ho, P.~T.~P.,
  Reid, M.~J., \& Schneps, M.~H.\ 1985, \apj, 293, 522

}
\end{thebibliography}
\end{document}